# Lecture I: Governing the Algorithmic City

Seth Lazar

Machine Intelligence and Normative Theory Lab

Australian National University[*]

## Abstract

A century ago, John Dewey observed that '[s]team and electricity have done more to alter the conditions under which men associate together than all the agencies which affected human relationships before our time'. In the last few decades, computing technologies have had a similar effect. Political philosophy's central task is to help us decide how to live together, by analysing our social relations, diagnosing their failings, and articulating ideals to guide their revision. But these profound social changes have left scarcely a dent in the model of social relations that (analytical) political philosophers assume. This essay aims to reverse that trend. It first builds a model of our novel social relations—as they are now, and as they are likely to evolve—and then explores how those differences affect our theories of how to live together. I introduce the 'Algorithmic City'—the network of algorithmically-mediated social relations—then characterise the intermediary power by which it is governed. I show how algorithmic governance raises new challenges for political philosophy concerning the justification of authority, the foundations of procedural legitimacy, and the possibility of justificatory neutrality.

[*] This essay, based on my 2023 Tanner Lecture on AI and Human Values at Stanford, will be published in my forthcoming book, *Connected by Code: How AI Structures, and Governs, the Ways we Relate*, with Oxford University Press, alongside commentaries by Joshua Cohen, Marion Fourcade, Renée Jorgensen and Arvind Narayanan, and my response to those commentaries. Please refer where possible to the published version. See the end for acknowledgments.

> *We are entering an age when the power of regulation will be relocated to a structure whose properties and possibilities are fundamentally different. … one form of power may be destroyed, but another is taking its place. Our aim must be to understand this power and to ask whether it is properly exercised. (*Lessig, 2006: 79*).*

## 1. INTRODUCTION

Political philosophy's central task is to help us determine how to live together. It analyses our social relations, diagnoses their failings, and articulates ideals to guide their revision. A century ago, John Dewey argued that '[s]team and electricity have done more to alter the conditions under which [people] associate together than all the agencies which affected human relationships before our time',[2] and called for political philosophy to be revised in light of these changes. In recent decades, computing technologies have had no less significant an impact on how we associate together than did steam and electricity. And political philosophy is overdue another update.[3] Our social relations are now infused with, sometimes even constituted by, computational systems. These may not enable, as Lessig thought, entirely new modalities of power. But the prevailing means of governing power have undoubtedly shifted.[4] And political philosophers must determine whether and how that power can be properly exercised.

Lecture I begins that task. The first step is to clarify how social relations have changed in the information age. I introduce 'the Algorithmic City'—the network of algorithmically mediated social relations of which we are now all part. I then characterise how power operates through these algorithmic intermediaries, highlighting a shift in both the means and prevailing modalities of power compared to our pre-algorithmic social relations. I argue that algorithmic governing power should not simply be eliminated but can be properly exercised—provided it aims at substantively justified ends and is used according to legitimate procedures by those with authority to do so. I then show how algorithmic governance raises presumptive challenges for familiar political philosophical approaches to these three normative standards. In the second Lecture, I use the approach developed at a high level here to consider a particular dimension of the Algorithmic City: the digital public sphere.

## 2. THE ALGORITHMIC CITY

John Rawls argued that we cannot select principles of justice for a society if we lack understanding of how that society works.[5] To make normative claims about how our social relations should be governed, we must understand those social relations' nature. Yet while some dimensions of our social relations—for example, sex, gender, race, nationality, and climate—have progressively garnered more of our philosophical attention,[6] another set of fundamental social changes has been too much ignored. 'The information age' describes the changes wrought by information and communication technologies grounded in computing. Political philosophy urgently needs updating for the information age. The task ahead is mammoth. This Lecture contributes by identifying and analysing what I consider the most significant political change computing has made: the advent of the network of algorithmically-mediated social relations, the Algorithmic City.[7]

---

[2] Dewey, 2016 (1926): 169.
[3] There are of course exceptions; for introductions to the political philosophy of AI, see Susskind, 2018; Coeckelbergh, 2022; Risse, 2023. For the first special journal issue on the topic, see Zimmermann et al., 2022.
[4] As I will explain below, the 'means' of governing power is the particular technology or practice used to govern—e.g. law, algorithms, parental injunctions. The 'modality' of governing power is the way in which that means is used to govern—e.g. by pre-empting, nudging, coercing, forcing.
[5] Rawls, 1999.

Social relations are stable patterns of communication and interaction between people.[8] This includes deeper connections, such as those between family members and friends; weaker ties, such as between colleagues, co-citizens, or participants in economic exchange; and even the thin connection of indirectly communicating through contribution to the same public discourse over time. I am most interested in social relations that are significant in people's lives, either individually or in the aggregate.

In the information age, our social relations are, to a growing degree, partly or wholly constituted by *algorithmic intermediaries*. These are the computational systems by which we now connect—social media, e-commerce, search engines, generative AI, email and messaging, and so on. I call them *algorithmic* intermediaries in the spirit of recent work in sociology and communication studies, which uses the term as a synecdoche.[9] Algorithms are programmed instructions executed by a computational system—either direct instructions detailing exhaustively what the system should do under different conditions or indirect instructions to the system to learn from a set of training data and then (in effect) write its own direct instructions. Algorithms are the tip of a long spear, part of a broader sociotechnical assemblage that includes other software and hardware elements of computational systems, human labour and material resources.[10]

Artificial Intelligence (AI) picks out a suite of algorithmic approaches to enabling computational systems to represent their environment and then act on it to achieve an intended outcome.[11] These approaches are functionally unified by this task description but are highly heterogeneous. Machine Learning (ML) algorithms are undoubtedly the ascendant force in AI, with transformer-based[12] self-supervised neural networks having enabled remarkable progress in Natural Language Processing (NLP), text generation, image recognition, image generation, and even agentic behaviour.[13] But other AI techniques—planning, knowledge representation, and logic, among others—are also highly societally influential. And other non-AI algorithmic methods play a prominent role in the Algorithmic City, such as algorithmic mechanism design, blockchain, hash-matching, and some non-ML varieties of information retrieval. Most of the algorithmic intermediaries that I focus on make prominent use of AI, but my analysis extends to other algorithmic systems that would not typically be considered in that (slightly amorphous) category.

---

[6] See e.g. Beitz, 1979; Okin, 1989; Mills, 1997.
[7] For other illuminating work on algorithmic intermediaries see e.g. Kohl, 2012; Langley and Leyshon, 2017; Elkin-Koren and Perel, 2019.
[8] To be clear, social relations can also obtain between a person and a company, or a person and a bot; but in each case they are (thus far) reducible to social relations between people— those who steer the company, or design and/or deploy the bot. Generative AI Systems may in future upset this simple reduction, but we are not there yet.
[9] See, for example, Gillespie, 2014; Bucher, 2018; Burrell and Fourcade, 2021.
[10] Bucher, 2018.
[11] See e.g. Russell and Norvig, 2016.
[12] 'Transformers' are an element in neural network architecture that allows ML models to faithfully represent not only the individual 'tokens' of the inputs they are modelling, but their relation to other tokens in a sequence. Vaswani et al., 2017.

Why call these *algorithmic* intermediaries rather than *digital* intermediaries? Here I am using 'algorithmic' metonymically, to pick out the feature of these sociotechnical systems that is most salient. Their crucial feature is that they are dynamic and adaptive, able to monitor the social relations that they mediate and to intervene in them in real-time. In other words, the algorithmic is the *agential* dimension of computing—the functional ability of computational systems to perceive changes in their environment and act on them. This (functionally) agential dimension of algorithmic intermediaries explains their ability to operate at such extraordinary speed and scale. The algorithmic component of these computational systems accounts for the transformative impact they have already had on our economies, our culture, our politics and our personal relationships, as well as the still greater changes that they portend. It also proves crucial to understanding whether and how the power relations they enable can be justified.

Intermediaries are go-betweens. They mediate between two or more mediatees, conveying information or action from one to another. Social relations are constituted by mutual communication and interaction.[14] If and to the extent that an intermediary conveys communication and interaction, then the intermediary at least part constitutes those social relations.[15] Intermediaries can be passive conduits; algorithmic intermediaries, however, actively shape the social relations that they constitute.

The early days of the internet saw many scholars explore urban and other metaphors at length; metaphors in general can be fertile but also misleading. I use the concept of 'the Algorithmic City' *strictly* as shorthand for 'the network of algorithmically mediated social relations'.[16] This network is not a *space*—like 'cyberspace'—supposedly distinct from the physical, non-algorithmic world. Algorithmic intermediaries infuse almost all our social relations, including those with substantial non-algorithmic dimensions. And the Algorithmic City describes how humans, in all our visceral, physical reality, connect to one another (and sometimes to bots). The algorithmic intermediaries connecting us are just software tools implemented on physical infrastructure—from servers and GPUs to cables and broadcast towers.

---

[13] At time of writing, the field has still not settled on a particular terminology for these systems. I will therefore adopt one set of definitions, acknowledging that others may become more current. Large language models (LLMs) are mathematical models of natural language, trained using deep, self-supervised learning to predict the next token in a string of tokens (where tokens are word-parts). Some LLMs are also trained on image data, and so have been called Large Multi-Modal Models. I won't make this distinction; they're all LLMs for our purposes—they all use autoregressive transformers to predict or generate sequences of tokens. We can distinguish between a pre-trained LLM, which has undergone only this training, and a 'fine-tuned' LLM, which has also been trained for some specific task through the use of supervised learning on labelled data (for example, a dataset of helpful responses to user questions), and/or with reinforcement learning with human or computational feedback. Fine-tuned LLMs are then deployed as part of more complex sociotechnical systems like ChatGPT—a dialogue agent created by OpenAI. I'll call these 'Generative AI Systems', even though the term 'generative' is a bit confusing here—many understand it to simply mean 'AI systems that generate content', rather than its technical meaning, which distinguishes between generative and discriminative classifiers (Ng and Jordan, 2001: 1). I will use 'Language Model Agents' to refer to the subset of Generative AI Systems that are designed to use other tools (besides just generating content) to achieve their ends.
[14] Benkler, 2006: 369.
[15] '[I]t depends on platforms not only bringing independent parties together but completely structuring every aspect of the exchange.' (Gillespie, 2018: 22).

The Algorithmic City is also different from the 'Society of Algorithms'—the latter concept aims to encompass every way algorithmic systems infuse society at large.[17] Political philosophy must urgently address the full gamut of algorithmic impacts on our social lives; but the Algorithmic City focuses on one specific dimension: the network of social relations mediated by algorithmic intermediaries. The 'Network Society' is an obvious ancestor to the Algorithmic City.[18] However, that concept places too much emphasis on the nodes of the network, and not enough on the edges (the connections between the nodes), which it presents as passive or neutral conduits. In the *Algorithmic* City the edges of the network are themselves able to dynamically update in order to reshape the social relations that they mediate and constitute. Although some algorithmic intermediaries can fairly be described as content-independent 'pipes', I am primarily interested in intermediaries that dynamically adapt and reshape the social relations that they mediate.[19]

Prominent examples of algorithmic intermediaries in the wild are: social networking sites, such as Facebook, Instagram, X (formerly known as Twitter), TikTok, YouTube, and Mastodon;[20] e-commerce sites, like Amazon and eBay; other two-sided markets for services, like Uber and AirBnB;[21] cultural two-sided markets like Spotify and Apple Music. We must also consider search engines,[22] generative AI systems like ChatGPT, Claude, and Gemini, app stores, operating systems, and even digital infrastructure like cloud and security services (e.g. AWS, Microsoft Azure, Google Cloud, Cloudflare, CrowdStrike).[23] Even mundane technologies like

---

[16] Thanks to Helen Nissenbaum and James Grimmelmann for discussion here.
[17] Burrell and Fourcade, 2021.
[18] Castells, 1996; Benkler, 2006.
[19] Drawing on the work of Bruno Latour, some argue that the concept of an 'intermediary' implies passivity and content-independence, using the word 'mediator' to mean what I describe as an intermediary. I reject this usage, which lacks (in English at least) linguistic or broader conceptual foundations. The concept of a 'mediator' in English typically refers to an intermediary mediating between parties that are at odds. Intermediary is the more general concept, and the more apposite for my purposes here, notwithstanding the Latourian associations. See Latour, 2005. For an example in this context, see Gillespie, 2017. Thanks to Gideon Futerman for discussion here.
[20] Van Dijck, 2013. Whether Twitter, TikTok, and YouTube are social networking sites is a fraught question, but it really does not matter here, since whatever else they are they are clearly algorithmic intermediaries. See e.g. Kwak et al., 2010.
[21] Birch, 2020; Viljoen et al., 2021.
[22] Introna and Nissenbaum, 2000; Noble, 2018.
[23] This list is intended to trace down through the layers of the software 'stack'. José van Dijck

email and messaging apps are algorithmic intermediaries—consider, for example, how email is automatically scanned and filtered, and can be monitored for different forms of non-compliance. But algorithmic intermediaries also encompass some of the most advanced developments in AI.

For example, the most prominent algorithmic intermediaries in our lives today are the digital platforms that determine what we see and engage with online. They in turn use many different computational means to shape how we connect to one another and to the internet's information, including relatively simple principles of interface design alongside recommender systems that deploy cutting-edge research in deep reinforcement learning to model the vast space of possible user-content combinations and determine which are most likely to be relevant to a particular user. In recent years, both these recommender systems and the search algorithms that shape our experience when we actively pursue content or information online have relied increasingly on fundamental progress in the design and deployment of large language models (LLMs). As noted in the prologue, the gap between fundamental research breakthroughs and deployment to millions, even billions of users, has narrowed to the point where basic research is being deployed within weeks of discovery.

LLMs use self-supervised deep learning on vast corpuses of text, images, and other data to create inordinately complex models of that data. These 'pre-trained' models are then fine-tuned for specific functions using supervised learning on labelled datasets, and reinforcement learning with human or computational feedback.[24] The resulting Generative AI Systems are then used to provide better representations of online content for both recommender systems and search, as well as to provide a search interface grounded in natural language dialogue rather than conjuring the correct text string to find what you need.[25] And many, many other use cases are being actively pursued. Some of these will undoubtedly fall flat, but there are few domains where Generative AI Systems do not hold the promise of revolutionising our existing algorithmic intermediaries. At a minimum, they are very likely to substantially mediate our access to each other, for example when they are used to generate text and images by which we communicate, or to filter and summarise communications we receive from others.[26] And sufficiently capable models can provide the executive control centre for AI Agents, which are able to carry out complex sequences of actions in order to achieve particular goals, including using other software tools, such as browsing the internet, drawing on symbolic reasoning systems and knowledge databases, and even controlling robotic systems.[27] These Language Model Agents may ultimately provide computational foundations for *universal intermediaries* that mediate between users and all digital technologies, enabling anyone to use computers for anything they are able to do, without needing to know the specific instructions necessary to achieve that end, or indeed

---

has proposed departing from this way of conceptualising platforms, advocating instead an arboreal visualization. As illuminating as it is, the visual metaphor has unfortunate implications, implying for example that the primary function of platforms in the 'trunk' of the platformization tree is to connect users to the infrastructural 'roots' of the tree, as distinct from connecting them to one another, or to businesses (Van Dijck, 2021).

[24] Vaswani *et al.*, 2017; Bai et al., 2022; OpenAI, 2023a, 2023b.
[25] Sun et al., 2023.
[26] See especially projects like Microsoft 365's Copilot and Google's Sidekick.
[27] Yao et al., 2022; Schick et al., 2023.

even needing to antecedently know that those ends were feasible objects of choice.[28]

Generative AI systems might end up fundamentally restructuring the Algorithmic City. Alternatively, we might find that natural language is an awkward user interface, and that fundamental security problems limit our ability to rely on Language Model Agents to be our digital butlers. Much is yet to be seen. But while their transformative potential is impressive, they are simply the most prominent recent manifestation of a much deeper trend. They could represent the apotheosis of the form, but *wherever* algorithmic systems are mediating between people, enabling them to communicate or to act on one another at a distance in a way that can be dynamically influenced by the intermediary, we have a corner of the Algorithmic City, in which algorithmic intermediaries are reshaping the social relations that they partly constitute.

While online platforms are paradigmatic algorithmic intermediaries, not all algorithmic intermediaries are online platforms.[29] Online platform companies provide a structure on which other products can be built, typically mediating between an audience and a set of companies providing services.[30] They are often two-sided markets, relying on algorithmic mechanism design to connect users with producers or service-providers.[31] By contrast, some messaging apps, like Signal, strictly intermediate between users, not between users and companies (or between users via companies). And some apps, like Tinder, algorithmically match users, rather than connecting users and companies. Mechanical Turk is arguably a platform, but other algorithmic intermediaries, like the software used to monitor performance of Amazon warehouse workers, are arguably not.[32] Surveillance tools—for example, apps developed to apply facial and object recognition to existing CCTV streams—are also algorithmic intermediaries, but are not platforms.[33] While the Algorithmic City is undoubtedly very online, algorithmic surveillance and management tools, as well as the prospect of locally deploying Generative AI Systems, show that not all algorithmic intermediaries live on the internet. In fact, just as the existing online platforms are exploring every possible way of integrating LLMs into their infrastructure, we are also seeing early signs of AI causing an interesting shift in those platforms' role. Smaller companies developing the most capable pre-trained models (OpenAI, Anthropic) are making deals with the online platforms that also have very large cloud data and compute services (e.g. Amazon Web Services, Google Cloud, Microsoft Azure). The AI companies then provide access to their pre-trained models to businesses, to be fine-tuned for specific purposes, often on proprietary data. These pre-trained models, often called 'foundation models', may be the foundation for a new twist to the

---

[28] See in particular platforms like LangChain and HuggingFace Transformers Agent, companies like Adept.AI, and initiatives like Google's Project Astra.
[29] On platforms, see Gillespie, 2018; Van Dijck et al., 2018; Van Dijck, 2021.
[30] Van Dijck *et al.*, 2018; Giblin and Doctorow, 2022.
[31] Viljoen *et al.*, 2021.
[32] Negrón, 2021.
[33] In the strict sense, an 'algorithmic intermediary' is the full assemblage of companies and products that mediate between two people, constituting (in whole or in part) the social relation between them. However, again taking part for whole, I will describe individual apps as algorithmic intermediaries, rather than 'an element of an algorithmic intermediary'.

platform economy.[34]

Algorithmic intermediaries are often associated with 'Big Tech' companies. Google's market capitalisation is founded on both its search algorithms and the algorithmic systems used to collect users' data and use it to predict click-through rates in its real-time auctions for ad placement. Meta, similarly, has both driven and profited from fundamental research in ML in developing its algorithmically-structured Feed for filtering and promoting content, and in its practices of surveillance advertising. As noted in the prologue, Big Tech and AI are symbiotes: the profits and power of Big Tech have enabled collection and curation of the vast datasets on which AI systems are trained, as well as the fundamental progress in computational capacity necessary to train very large AI models; and AI has enabled the Big Tech companies to achieve and grow their market power.

And yet, it would be a mistake to conflate our analysis of the Algorithmic City with a critique of the practices of Big Tech companies, for both methodological and empirical reasons.

On the methodological commitment: as noted in the Prologue, the role of philosophers in a fast-moving real-world debate must be to try to identify the underlying nature of things, the forms beneath the appearances. Even over the course of writing this book, the fortunes of the Big Tech companies, as well as the composition of that cohort, have altered radically. These febrile vicissitudes are a distraction from the task of posing and answering the deeper political philosophical questions that algorithmic intermediaries raise.

And descriptively, not all AI advances deployed by Big Tech constitute algorithmic intermediaries—consider, for example, the extraordinary progress in computational photography over the last decade. And while Big Tech strives to own a piece of any successful algorithmic intermediaries, not all algorithmic intermediaries are run by Big Tech.

First, many algorithmic intermediaries are products of 'Little Tech'. For example, in a recent report Wilneida Negròn catalogues the many smaller companies developing algorithmic management tools.[35] And we are in the middle of a 'Cambrian Explosion' of AI start-ups, aiming to use Generative AI to create new kinds of algorithmic intermediaries (though by the time this book is out, that explosion might have transformed into an implosion as the more successful companies are absorbed into Big Tech counterparts). Perhaps most importantly, algorithmic intermediaries developed by huge finance companies play an extraordinary role in the global economy. For example, BlackRock's Aladdin investment management and trading platform involves risk analytics, trading and operations tools. It is the quintessential algorithmic intermediary, and it manages tens of trillions of dollars of assets. It would be omitted from an analysis that focused only on 'Big Tech'.

Second, algorithmic intermediaries include much more than just contemporary centralised for-profit businesses. They also include decentralised for-profit architectures like those associated with cryptocurrency and blockchain, as well as non-profit/public-benefit counterparts like the 'fediverse' of federated, decentralised internet services, and even centralised, state-controlled algorithmic intermediaries like those answerable to China's techno-authoritarian government. And the model also applies to future developments in digital social relations—including visions from the internet as diverse as those associated with web3, the 'metaverse', universal intermediaries, and algorithmically-mediated democracy.[36]

---

[34] Bommasani et al., 2021.
[35] E.g. Negrón, 2021.

Algorithmic intermediaries play such a pervasive role in our lives that it might seem difficult to identify the boundaries of the Algorithmic City. But not all AI systems are algorithmic intermediaries, because not all AI systems mediate communication and interaction. Most notably, one of AI's most prominent social roles is for *analysis*, to support human decision-making that is not algorithmically-mediated. For example, judges use ML models to inform their judgments of whether to grant bail or parole.[37] Government bureaucracies use ML to analyse individual and societal behaviours to decide how to allocate resources, and where to target interventions.[38] ML models of the prospective path of an infectious disease inform public health judgments, and ML models might flag individual children as being at risk of abuse, leading to a visit from care workers.[39] None of these are examples of algorithmic intermediaries.

## 3. THE NATURE AND JUSTIFICATION OF ALGORITHMIC POWER

### 3.1. Dimensions of Algorithmic Power

The Algorithmic City is a model of how digital technologies transform our social relations. A model's worth comes down to its explanatory utility. Drawing attention to algorithmic intermediaries foregrounds adapted and intensified power relations that urgently demand normative evaluation.

Power is, roughly, the ability to shape others' prospects, options, and attitudes (beliefs and desires).[40] A has power over B just in case A can exercise power over B, so defined. One of political philosophy's central questions is, in Lessig's terms, when is power properly exercised? So political philosophers should closely attend to the ways in which new technologies introduce new or newly intense power relations, or alter the distribution of power. The first step in this process is to empirically identify the variety of ways in which the new technology affects the social distribution of power. Second, we need to analyse these new distributions of power, to identify whether the new technology raises new kinds of questions about power's justification. And third, we must determine whether and how this new distribution of power can be justified.

I want to start by drawing attention to three dimensions of how algorithmic intermediaries shape social power relations. First, algorithmic intermediaries enable those who design or deploy them to exercise *power over* the people whose social relationships the algorithmic intermediaries mediate. Second, algorithmic intermediaries shape power relations *between* the mediatees. Third, through exercising power over us, and reshaping power relations within our social relations, over time algorithmic intermediaries reshape our broader social structures. I'll elaborate on each in turn.

---

[36] See e.g. Siddarth et al., 2021.
[37] Angwin et al., 2016.
[38] Engstrom et al., 2020.
[39] Schwartz et al., 2017; Stapleton et al., 2022.
[40] I draw here on my analysis of power in Lazar, 2021, which itself draws heavily on the debate occasioned by Dahl, 1957; Bachrach and Baratz, 1962; Lukes, 2005. Although I came upon it after writing this work, Grimmelmann, 2015 on moderation is very much complementary.

First, power over. Algorithmic intermediaries do not spring up from the ground like mushrooms. They are complex sociotechnical systems designed and deployed by self-interested actors in the (often disappointed!) hopes to achieve their own goals. As such, they enable those behind these systems to exercise power over those whose social relations the systems mediate. To say that an algorithmic intermediary exercises power over its mediatees is, in general, just to say this: the intermediary enables those who design or deploy the system to exercise that power. Since this is the most common case, this is what I will mean when I say that 'algorithmic intermediaries exercise power over their mediatees'. I consider exceptions to this rule—where the algorithmic intermediary *itself* exercises power—in the next subsection.

Algorithmic intermediaries exercise power over their mediatees by directly affecting their prospects—making them better or worse off. This might come through, for example, amplifying or demoting your communications, suggesting and enabling new connections, or banning and removing you from a platform. Search algorithms in two-sided markets can make or break a business.[41] Algorithmic management tools can make a worker's life intolerable.[42] Given that algorithmic intermediaries can in-principle access and record every aspect of the social relations they mediate, they also have unprecedented enforcement powers over mediatees.[43] Scholars since the 1990s have recognised the quasi-legal power of code—what Reidenberg called *Lex Informatica*.[44] Algorithmic *intermediaries* add another layer: they shape our access to *other people*. Exclusion by the intermediary could lock you out of a vital market, or disrupt a meaningful relationship. Algorithmic intermediaries' power derives not only from the disciplining dimension of code but from how they can enable or limit access to the people with whom we wish to interact.[45]

By constituting the relationships they mediate, algorithmic intermediaries substantially determine our options in those relationships.[46] Everything we can communicate or do to one another via an algorithmic intermediary is made possible by that intermediary. This *productive* power is pervasive in digital environments, often enabling new kinds of action. Generative AI Systems provide perhaps the most striking example of this in recent years, as they enable people to communicate with others in ways that they were literally incapable of just months ago; Language Model Agents promise to do the same for many more domains of human action besides simply generating text and images. Algorithmic intermediaries also remove options, making certain behaviours impossible for those whose relationships they mediate (at least, through that intermediary).[47] Digital 'locks' like Digital Rights Management, take this form.[48] Meta's personal boundary in its Horizons virtual world is another example. Rather than mandating that users ought not to grope one another's avatars, the algorithmic intermediary simply removes the option of invading another avatar's personal space.[49] And efforts by leading AI research companies to 'align' their Generative AI Systems to 'human values' are most often further examples of just this kind of option-elimination, as people are prevented from using the models to generate certain kinds of content, or performing certain kinds of actions. These options are stripped away, sometimes, by filters that prevent either certain kinds of instructions from being acted on, or certain outputs being generated; even more interestingly, they also train the models to recognise potentially harmful instructions, and steer them towards safer ground.[50]

---

[41] See e.g. Khan, 2016; Romm et al., 2020.
[42] Negrón, 2021.
[43] On 'perfect enforcement', see especially Zittrain, 2008.
[44] Reidenberg, 1997.
[45] See e.g. Giblin and Doctorow, 2022.
[46] 'M[ulti ]S[ided ]P[latform]s are not neutral intermediaries. Arguably, they are not even really markets. Instead, they are tools and enactments of world building: their designs and optimization strategies channel power, structure forms of decision making, and encode economic and legal relationships.' Viljoen *et al.*, 2021: 9.

All technologies afford some options and push against others.[51] Algorithmic intermediaries are particularly well adapted to shape our choice architectures. Simple design choices, such as what kinds of response a platform allows to another person's communications, as well as more intentional 'dark patterns' nudging us towards particular outcomes, make some options for communication more appealing than others, raising barriers to disfavoured behaviour, and encouraging favoured behaviour.[52] In the extreme, this includes adding penalties to some options (and incentives to others). This approach is typically used when pre-emptively removing disfavoured options is undesirable.[53]

And algorithmic intermediaries shape our beliefs and desires. We learn about the world through recommender systems, search engines, and now LLM-enabled tools like ChatGPT and Perplexity.[54] Algorithmic advertising and the dependence of our information environment on recommender systems anticipate and cultivate our current appetites, and generate new ones, such as the opportunity to win a minimal form of approbation from strangers on the internet (a like, upvote, or repost).[55] This trend will radically accelerate as Language Model Agents enable universal intermediaries to digital technologies—instead of simply receiving recommendations in the form of a ranked list, we will routinely interact more with AI-generated summaries (including selection and editorialising) of other people's

---

[47] Lessig, 2006. Brownsword has written extensively on this theme; for a comprehensive overview of his approach, see Brownsword, 2022. For further germinal work in this vein, see Hildebrandt, 2015.
[48] Kerr, 2009.
[49] https://www.businessinsider.com/meta-metaverse-virtual-groping-personal-boundary-safety-bubble-horizons-venues-2022-2
[50] Bai *et al.*, 2022; OpenAI, 2023b.
[51] Gibson, 1977. See also Winner, 1980. For a recent and comprehensive overview of this problem and literature, see Davis, 2020.
[52] Brock, 2012; Yeung, 2017; Vaidhyanathan, 2018.
[53] Brownsword, 2022, calls this 'governance by machine' as distinct from technological management. Yeung, 2018, describes it as one modality of 'algorithmic regulation'. Lessig, 2006, also distinguishes between coding problems away and more effective enforcement.
[54] Introna and Nissenbaum, 2000; Noble, 2018.
[55] Recommender systems based on reinforcement learning will also actively try to shape our preferences to make them easier to satisfy Carroll et al., 2022.

speech than with those people themselves.

Second, power between. Critics often think of digital technologies as simply dominating—vehicles for the arbitrary power of Big Tech companies, digital thumbs pressing us down. But while that is clearly happening too, as just described, many (perhaps most) digital pathologies are caused by other people— who can communicate with and act on us because of the algorithmic intermediaries mediating our social relations. As well as enabling those who design and deploy them to exercise power over those they mediate between, algorithmic intermediaries also shape power relations between mediatees, enabling some of them to exercise power over others.

Algorithmic intermediaries enable some to directly affect others' prospects. For example, as I discuss in Lecture II of this book, in the digital public sphere algorithmic intermediaries can incentivise forms of individual and collective harassment and abuse that very clearly make their victims much worse off, sometimes causally contributing to serious material harms, like suicide and even genocide.[56] Algorithmic intermediaries also empower some to shape others' options, by enabling various kinds of economic, social, and cultural interactions— and by determining which are encouraged and which discouraged. And they allow some to shape others' beliefs and desires. Commentators lament how 'the algorithm' is leading people astray, manipulating them, filling their heads with misinformation, and so on.[57] But while recommender systems undoubtedly play a role, the bottom line is that *a person* (or bot) is producing manipulative content, and then, through algorithmic intermediaries, manipulating others. Whether or not 'the algorithm' is manipulating people, algorithmic intermediaries have created great incentives and opportunities for people to deceive and manipulate one another.[58]

Like any city, the Algorithmic City is fundamentally a site for interaction between people. It enables us to connect to one another—to forge new social relations of commerce, culture, sociality and politics. And any medium that enables interaction and communication will either help some people to have unjustified power over other people, or else support egalitarian social relations. People's propensity to dominate one another, to use the means available to them to seize resources and power, is parametric. The means we have for communicating and acting on one another will shape the degree to which we are able to realise those goals, thereby shaping power relations between the mediatees. This will ultimately prove essential to the evaluation of algorithmic power.

Third, power through. Algorithmic intermediaries shape our social structures over time by shaping the social relations they mediate. Social structures are, roughly, networks of roles, relationships, incentives, norms, cultural schemas (widely shared sets of evaluative and doxastic attitudes), and institutions, which can be populated or observed by different people at different times; they are generally the emergent result of patterns of human interaction over time, and they reliably pattern outcomes for people who are within or otherwise affected by them.[59] They are partly constituted by social relations, so if algorithmic intermediaries change our social relations, then, over time, they also change the social structures they constitute. In general, social structures are hard to change; they are particularly hard to change intentionally. Dewey beautifully captures this idea:

> In spite of sudden and catastrophic revolutions, the essential continuity of history is doubly guaranteed. Not only are personal desire and belief functions of habit and custom, but the objective conditions which provide the resources and tools of action, together with its limitations, obstructions and traps, are precipitates of the past, perpetuating, willy-nilly, its hold and power.[60]

---

[56] Siegel, 2020.
[57] Some recent examples: Barack Obama, Jon Stewart.
[58] Munger and Phillips, 2020.
[59] See e.g. Haslanger, 2016; Ritchie, 2020.

Algorithmic intermediaries are arguably unique for their impact on these 'precipitates of the past', across all major spheres of human interaction—notably politics, economics, culture and (affective) sociality. Our public sphere and political discourse have been remade.[61] Analogue markets structured by laws and regulations are increasingly being replaced by algorithmic quasi-markets, structured primarily by their designers' intentions and interests.[62] Algorithmic intermediaries have enabled companies to generate mass audiences for cultural products, channelling creators through 'chokepoints' where they are forced to accept punitive terms to reach that audience.[63] Even friendship and dating have been reshaped.[64] Generative AI Systems have *already* radically changed the nature of education, overnight constraining the deployment of familiar and inclusive techniques of student assessment (as well as offering the promise of a new kind of instruction). And with the advent of AI companions that simulate many features of human agency, they are set to have further transformative impacts on the nature of sociality and perhaps even sexuality.

These social transformations never arise in a simple 1:1 relationship to the designers' intent. Results are often unintended and reflect a complex dynamic of top-down influence and bottom-up resistance and appropriation.[65] But still, algorithmic intermediaries have both changed our occurrent social relations and, by doing so, radically changed the social structures that trammel our lives.[66] The sheer speed at which they have operated demands our attention: past technological advances have taken years, decades, even centuries to go from fundamental research breakthrough to significantly altered social structures. Algorithmic intermediaries can be instantly deployed to billions of people around the world. Their societal impacts therefore proceed at a radically accelerated pace.

3.2. Analysing Algorithmic Power

Whenever new technologies enable new or newly intensified power relations to emerge, or otherwise shift the distribution of power in our social relations, political philosophers should take notice. And we face a dual challenge: first to come to understand the nature of these new distributions of power, and then to morally evaluate them.

---

[60] Dewey, 2016 (1926): 186.
[61] Gillespie, 2018; Vaidhyanathan, 2018; Suzor, 2019; Andrejevic, 2020; Persily and Tucker, 2020. More generally, see Lecture II.
[62] E.g. Cohen, 2019; Birch, 2020; Viljoen *et al.*, 2021.
[63] Giblin and Doctorow, 2022.
[64] Van Dijck, 2013; Bucher, 2018.
[65] See e.g. Van Dijck, 2013; Bucher, 2018; Gillespie, 2018: 32.
[66] 'Sociality is not simply "rendered technological" by moving to an online space; rather, coded structures are profoundly altering the nature of our connections, creations, and interactions.' (Van Dijck, 2013: 20).

To attempt the former task, it will help to distinguish between *means* and *modalities* of power. A means of power is a tool, practice, or technology that enables the exercise of power. The law, for example, is a canonical means of power. But so is the parental voice, so are the competitive pressures of a free market, or the physical infrastructure of an analogue city, or the extra-legal norms and conventions that shape human behaviour. *Modalities* of power are the ways in which a particular means of power can be used to exercise power. In the foregoing section, I distinguished between exercising power by directly affecting people's prospects, by shaping their options, and by shaping their beliefs and desires. These are all different modalities of power, at one level of description. In this subsection, I introduce further modalities of power that cut across those.

A priori, any means of power can be used according to any modality of power. Nonetheless, different means are likely to favour different modalities, just as in general the nature of technologies shapes the functions that they can successfully be used to perform. Algorithmic intermediaries are an historically novel *means* of exercising power. And—like all means of exercising power—they favour a distinctive combination of modalities of power, which demands careful evaluation in its own right. Intriguingly, the advent of new means for exercising power offers the prospect of a kind of natural experiment in understanding the modalities of power—by coming to understand the nature and justification of the new mix of modalities, we can learn more about our other means of exercising power, and perhaps shine a light on modalities of power that have previously received inadequate attention.

To support this point, I want to introduce some further distinctions in the modality of power. The most important is the distinction between *extrinsic* and *intermediary* power.

Extrinsic power shapes social relations from the outside in: it creates physical spaces and institutional parameters within which people can engage in unmediated interaction. Our analogue cities involve much extrinsic power. Physical spaces shape which kinds of interaction are feasible, and laws determine boundaries for permissible behaviour. Compliance with those laws is up to you—though choosing non-compliance risks penalty. This extrinsic structure permits participants in social relations to interact and communicate in an agentially-unmediated way. To adapt a metaphor from Dewey: extrinsic power governs social relations the way a river's banks govern the water's flow.[67]

Intermediary power is more like the bonds knitting the water's molecules together. It constitutes (in whole or part) the social relations that it mediates, thereby determining what shape they can take—what is possible or impossible, encouraged or frustrated. It progressively eliminates the possibility of agentially-unmediated communication and reduces the scope of feasible non-compliance, since refusal is impossible unless it is intentionally designed in. Intermediary power shapes social relations from the inside out.

---

[67] Dewey, 2016 (1926): 99.

Algorithmic intermediaries, like most means of power, can operate both extrinsically and instrumentally. When algorithmic intermediaries are used by the agents of extrinsic power to surveil populations, monitor them for compliance with some set of rules, and subject them to penalties for non-compliance, they are implicated in the exercise of extrinsic power. However, as the name suggests, algorithmic intermediaries also make extensive use of intermediary power. Indeed, in one sense, they bring it to an unprecedented apotheosis.

To be clear, intermediary power is not new to the Algorithmic City. This modality of power has always been present, just as there have always been intermediaries, such as negotiators, bureaucracies, brokers, curators, critics, media companies and matchmakers. In literature, we have characters like Iago from Shakespeare's *Othello*, or the Lawyer in Kafka's trial. They have always been able to exercise power over those they mediate between.[68] They monitor the behaviour of their mediatees, learn and shape their beliefs and desires, and then shape their relationship, typically in ways that profit the intermediary. Call this kind of adaptive, individualised intermediary power *agential* intermediary power.[69]

Other means of power, like the law, also manifest this modality of power. For example, the law often plays a constitutive role, in which it makes certain kinds of socially recognised options legally possible (for example, in how it defines marriage, or other types of contractual relationship, or the merger of two corporations).[70] Some would also argue that social norms exercise a kind of intermediary power, and more broadly that social structures such as economic systems, or even the language that we use, shape what is possible for people within their social relations, just as human or algorithmic intermediaries do, and therefore also exercise intermediary power. Language makes some intentions unsayable, some actions unthinkable, and directs us, through the terms we must use, towards some outcomes and away from others. Let's call this *structural* intermediary power, because it imbues social structures like laws, norms, and institutions, is never within the voluntary or unilateral control of any particular agent, and is hard to update (it is a 'precipitate of the past').

Algorithmic power stands apart for how it combines these two dimensions of intermediary power. It is both structural *and* agential. Algorithmic intermediaries' ability to operate at vast scale and lightning speed enables them to imbue social relations to a degree that is simply unmatched by any other agential intermediaries. They can shape social relations through shaping billions of microtransactions a second, as with the complex systems of real-time bidding on online advertising, and algorithmic stock trading. Online platforms' ability to shape our options, beliefs and desires at scale is truly structural—a change in Google's search algorithm will radically change economic opportunities within affected communities; if Facebook, Instagram, YouTube, TikTok, or X (Twitter as was) rejig their recommender systems they will affect how billions of people come to know about and understand the world; alterations to their real-time bidding systems for allocating advertisements online will reshape vast, multi-billion dollar industries. BlackRock's algorithmic investment management platform, Aladdin, manages *$22 trillion* of assets.[71] This is pretty clearly structural power! If Iago were able to be go-between for billions of distinct relationships simultaneously, then his already troubling power would be very central to the concerns of political philosophy, not just interpersonal ethics.

---

[68] On which, see for example Rosenfeld and Thomas, 2012; Bessy and Chauvin, 2013; Euchner, 2019; Young et al., 2019; Wang et al., 2022

[69] Note that algorithmic intermediaries may only simulate agency, or be functional agents; my analysis does not depend on their meeting any more internal or qualitative criteria for agency. It's enough that they can dynamically represent their environment and act within it to achieve a set of goals.

[70] Heverly, 2006. See in particular Hart, 1961 on secondary rules, and Rawls, 1955; Searle and Searle, 1969 on constitutive rules.

Algorithmic intermediaries therefore seem able to exercise structural intermediary power, pervading social relations at scale. Indeed, they are arguably even *more* pervasive than laws, norms, and institutions, since the latter tend to be confined to the borders of a nation-state, whereas algorithmic intermediaries can mediate social relations transnationally. But the remarkable feature of algorithmic intermediaries is how they combine this structural pervasiveness with an agential capacity to adapt, refine, and personalise their methods of control based narrowly on their environment and goals, in ways that are simply infeasible for analogue structural systems.

For example, dialogue agents based on LLMs have been trained over a vast amount of iterated simulations to interpret the intent behind users' questions, and give answers tailored directly to that intent, enabling them to never give the same answer twice. Recommender systems likewise respond adaptively to perceived facts about user identity and behaviour to adaptively shape how they connect with others online. This personalisation and fine-tuning proceeds in line with the commercial or political goals of the organisation that has designed and deployed the algorithmic intermediary; these are often hierarchical commercial or political organisations, whose leaders then have the unprecedented ability to unilaterally reshape how those algorithmic intermediaries exercise power, over billions of people at a time, without relying on anyone else's voluntary or intentional compliance.

Analogue structural intermediaries cannot match this ability for adaptive fine-tuning in real time. Laws operate best when they proceed via the will of those subject to them, so that we comply with the law voluntarily without needing to be forced. This means that in the absence of effectively comprehensive enforcement (which enables the will to be bypassed, more or less), significant legal changes in general need to be introduced gradually, with much public dissemination and preparation. Language is never in the voluntary control of any particular actor, but can only be changed over long periods of time through concerted efforts by many different parties. Algorithmic intermediaries that structure the social relations of billions of people can be changed at a single individual's whim—as when Mark Zuckerberg decided to alter Facebook's News Feed algorithm to reduce the spread of politically polarising communications.

This combination of structural and agential power is genuinely intriguing. Political philosophers have, for a long time, critiqued structural power in capitalist societies. This critique, however, routinely hits hurdles when it comes to articulating a theory of social change, and allocating responsibility for realising that change.⁷² This is unavoidable: social structures are in general hard to change. They typically emerge as and are sustained by largely unintended consequences of the uncoordinated behaviour of many different individuals and institutions. They are deeply embedded into many different social practices, which are typically objects of significant contention. If agreement could even be achieved on how to change them, doing so would involve resolving any number of coordination and collective action problems. And allocating either forward- or backward-looking responsibility for emergent social structures to any particular individuals or groups is very hard indeed. The same problems beset the more Foucauldian critique of the structural power inherent in language, for example.⁷³ Foucault identifies many ways in which social structures like language exercise dominating power, which we can infer is something undesirable, but he offers little explanation for *why* this structural power is bad, or argument for what we should do about it. This 'cryptonormativity' is arguably a rational response to purely structural power, for which even identifying appropriate agents of change is a challenge, let alone coming up with a tractable plan of action for how these structures could plausibly be changed.⁷⁴

---

[71] Coates and John, 2018. See also https://www.blackrock.com/aladdin.

Algorithmic intermediary power affords a notable contrast. It infuses social structures and can influence their long-term shape, but it is itself functionally agential, and enables a significant degree of fine-grained agential control. This means it is subject to a kind of direct normative evaluation that is less appropriate for social structures that lack this agential dimension. It also affords both a theory of change and an allocation of responsibility that are directly actionable. Algorithmic intermediaries are not emergent properties of complex sets of uncoordinated individual choices—they are discrete actors, which can exercise structural power at scale in relatively controlled ways. Of course, exercising power at significant scale is always a dicey and complex proposition—to say that algorithmic intermediaries exercise structural and agential power is not to say that they provide the secret strings to puppeteer the world. But algorithmic intermediaries enable dynamic, adaptive, targeted interventions and updates in response to information about the environment in which they are operating.

Their combination of structural and agential power is algorithmic intermediaries' most notable feature; I want to highlight one further observation about the nature of algorithmic power. Some algorithmic intermediaries are not merely tools that enable the exercise of agential intermediary power; they are *themselves* exercising power in the relevant sense. We can say that an AI system itself exercises power when it makes decisions that materially affect peoples' prospects, options and attitudes, in ways that are not antecedently foreseeable and controllable by those who either design or deploy the system. The most advanced AI systems are the best examples of this. It's important to be clear here. Existing algorithmic intermediaries are always *also* tools for the exercise of power by some over others. But they sometimes also exercise power themselves. These are not mutually exclusive possibilities. And saying that the AI system exercises power does not imply that it is morally responsible, or has any other morally relevant properties of agency. It simply means that the system is making decisions that are not substantially in the control of those who design or deploy it.

---

⁷² See especially Young, 2011 and the ensuing literature.
⁷³ Brown, 2008; Foucault, 1977, 2019.
⁷⁴ Kolodny, 1996.

Take Generative AI Systems based on LLMs like OpenAI's GPT-4o and Anthropic's Claude. As noted above, companies like OpenAI and Anthropic invest significant resources in 'aligning' these agents to human values. While they are not perfectly transparent about how this process works, a key element is reinforcement learning with computational feedback (what OpenAI calls 'rules-based reward modelling', and Anthropic calls 'Constitutional AI').[75] Once we understand how these—ingenious—methods work, two things become clear: aligning foundation models is really about exercising power over their users; and on this method of alignment the LLMs themselves are also exercising power, not only those who design or deploy them.

Here's how reinforcement learning with computational feedback works. First, supervised learning is used to train an LLM to be good at following instructions. Then, the instruction-tuned model is given two distinct tasks. One instance generates two completions each in response to many (many) prompts, and the other instance ranks those completions against a set of principles (what Anthropic calls Claude's 'constitution').[76] This feedback is encoded into a reward model, on which the instruction-tuned LLM is then further fine-tuned using reinforcement learning. In somewhat metaphorical terms, this effectively encodes the model's understanding of its (natural language) principles directly into its (mathematical) weights. This process is repeated (with new principles being added as necessary) until the fine-tuned LLM reliably generates 'value-aligned' responses to any given prompt (with a lot of human as well as computational evaluation). Importantly, with models as complex and intrinsically opaque as these, it is impossible to offer formal guarantees for their behaviour.

This—innovative and extremely philosophically interesting!—approach is part of the field of AI Safety, focused on value alignment. This is usually conceived as a way of governing LLMs. But this is slightly misleading. The real target of LLM-alignment is in fact the end user—the person who tries to use Claude to give them instructions on how to make a bomb, or to deploy GPT-4 in a Language Model Agent tasked with catfishing an unsuspecting third party. The AI companies aim to design their algorithmic intermediary so that it removes these options from the table, exercising quintessentially agential intermediary power, which can respond to subtle nuances and particularities in the environment and initial prompt. But it's not only the AI companies exercising power here. The models they have trained do so too.

The 'constitution' used by Anthropic, and one assumes the (secret) rules developed by OpenAI, are filled with redundancy, ambiguity, and conflicts. Claude's first constitution was a charming hodgepodge of concatenated principles from the Universal Declaration of Human Rights, Apple's Terms of Service, and injunctions to adopt a 'non-western' perspective.[77] Many of the clauses in the 'constitution' overlap or repeat one another; and they are all framed as being maximally important, without any hierarchy among them. In other words, *the model itself*

---

[75] Bai *et al*., 2022; OpenAI, 2023b.
[76] In fact, Constitutional AI involves ranking completions against one principle from the constitutional set.
[77] https://www.anthropic.com/index/claudes-constitution.

*performs a vast proportion of the task of interpreting and applying the constitution without any human input*. The LLM is also somewhat responsible for determining when the model is 'safe' for release, since the lack of formal guarantees means the only way to verify model performance is through vast AI-monitored simulations. As a result, it seems clear that when an LLM shapes its user's options, the LLM itself is exercising power, as well as those who design and deploy it.[78] Importantly, in the terms I will introduce properly below, LLMs like Claude and ChatGPT *govern* their users—insofar as the AI system itself determines, within boundaries, the norms that constrain that use.

### 3.3. Eliminate or Justify Algorithmic Intermediary Power?

Much normative work on the Algorithmic City has been reflexively critical, identifying moral shortcomings of existing technologies but rarely offering deep justification for *why* they constitute moral shortcomings, relying instead on implicit normative foundations, presumptively shared with the audience, assumed not to need support. On this view, naming power is enough to criticise it; distinguishing between permissible and impermissible power is anathema. An alternative approach moves directly from criticism to intervention, proposing regulations or technological solutions to mitigate these agreed-upon harms.

Both approaches are missing an important further step: the step where we reflect on and argue for the guiding principles that should shape power relations in the Algorithmic City. This is not the anaemic process of agreeing on 'AI Ethics Principles', but rather is about determining what we care about, so our moral diagnoses have depth and our regulatory or technical interventions have purpose. This is political philosophy's task (whether undertaken by those trained as philosophers or not). In the present context, its first objective is to consider the foregoing descriptive and interpretative account of how algorithmic intermediaries shift power relations, and explain whether their doing so is morally defensible.

The first step is to explain why the critics are right to be presumptively suspicious of new, and newly intensified, power relations. Presumptive hostility to power makes sense if you endorse one or all of three values which, for the purposes of this book, constitute the bedrock of my normative analysis of the Algorithmic City: individual liberty, relational equality, and collective self-determination. Together these form the bedrock of a liberal egalitarian democracy. Each value can be interpreted differently, and the ensuing discussion should be robust across most reasonable interpretations. However, for clarity, I will precisify them as follows.

I understand liberty as negative liberty or protection from wrongful interference *and the risk of wrongful interference* by others.[79] Negative liberty contrasts with *positive* liberty, which prioritises the ability to make authentic choices between desirable options, and *republican* liberty, which prioritises not minimising the risk

---

[78] Note that different companies may design and deploy these systems, and might therefore exercise power in subtly different ways. Thanks to Gideon Futerman for raising this important point.

[79] [Kramer, 2008](). Note that Kramer's view of negative liberty is morally neutral, whereas I think we should not view justified interference as a limitation on people's negative liberty. Nothing substantial in what follows should change if you prefer to adopt a non-moralised conception of negative liberty.

of interference but eliminating its possibility.[80]

The ideal of relational equality has deep roots, but it came to prominence in contrast with the philosophical focus on *distributive* equality.[81] Rather than focusing on the distribution of some good within a population, it describes an aspiration that we should live in a society where we recognise one another as moral equals, and where the institutions structuring our interactions reflect and support that equality.

Over time, societies collectively, and largely unintentionally, create and sustain social structures that affect our choices, making some things possible and others impossible, shaping our beliefs and desires. Collective self-determination is the process of reducing our subjection to heteronomous social structures that inadequately reflect our values. It involves jointly seizing the reins of our shared lives, so that we are not only formally equal, but we actually have positive political power to shape the shared terms of our social existence.

I call these *democratic* values because, in this world, democratic institutions are the only means by which all three will be realised. Moreover, democratic institutions constitutively enable relational equality and collective self-determination: they are not simply means to realise those values (as is I think the case for individual liberty), but rather democratic institutions themselves realise egalitarian social relations, and constitute collective self-determination. The institutions of democracy are many and complex, and I will not attempt a catalogue. And democracy is in practice always an imperfect ideal. But no other institutional arrangement has yet instrumentally or constitutively enabled the fulfilment of these foundational values. Democracy's enduring appeal is reflected in our appetite to defend it when it is threatened.[82]

With these three values in mind, we can see why new power relations warrant suspicion and how they must be justified or resisted. Very simply put: if A has power over B, then B is subject to the risk of wrongful interference by A. So, power is presumptively in tension with negative liberty.[83] If A has power *over* B, then they presumptively stand in hierarchical social relations, undermining relational equality. And if A has power over B, C, and D, then, presumptively, the society comprising [A, B, C, D] together are not collectively self-determining.

Algorithmic intermediaries exercise power over us. They are used to exercise power over us by those who develop and deploy them. They also shape power relations between those they mediate between—and by all accounts, they do so in ways that undermine liberty, equality, and collective self-determination (see Lecture II of this book for one argument to this effect). Should we therefore conclude that algorithmic power is straightforwardly morally indefensible? Is it simply something that should be eliminated, having no place in a liberal egalitarian democracy? Undoubtedly this is true for some of the leading figures in tech—their power (and fortunes) have no place in a just society. But the power of algorithmic intermediaries as such is different for three reasons.

---

[80] Goodin and Jackson, 2007; Pettit, 2008. I actually think positive liberty may prove central, in the end, to evaluating the Algorithmic City. But in the interests of being ecumenical I will focus on the thinner conception of negative liberty. My reasons for focusing on 'probabilistic negative liberty' as distinct from republican liberty concern both the impossibility and undesirability of eliminating the possibility of republican domination, and the tendency of the republican ideal of non-domination to conflate an internally diverse suite of normative concerns into a single value that has less explanatory power than separately focusing on (probabilistic negative) liberty, relational equality, and collective self-determination.
[81] Anderson, 1999; Scheffler, 2003.
[82] For related ideas, see for example Christiano, 2004; Viehoff, 2014.
[83] In fact, on the probabilistic understanding of negative liberty that I favour, power is necessarily in tension with negative liberty. However, even on other non-probabilistic accounts it is presumptively in tension. Thanks to Kyle van Oosterum for this clarification.

First, people will not forswear algorithmic intermediaries now or in the future—the benefits that they gain from using them are much too great. However harmful they can be, we lack sufficient grounds just to outlaw their development or use. The decision to connect with others through algorithmic intermediaries is utterly different from that to contract into slavery, for example. Algorithmic intermediaries are therefore inescapable features of our social landscape. Since they inevitably give their designers (and the systems themselves) power over those whose relationships they mediate, we must determine whether and how that power can be held appropriately.

Second, algorithmic intermediaries shape power relations between mediatees—like the law, they 'normally serve the primary purpose of setting the citizen's relations with other citizens'.[84] Dewey's description of the role of a good state also describes the task facing algorithmic intermediaries:

> [A good state] renders the desirable associations more solider [sic] and more coherent; indirectly it clarifies their aims and purges their activities. It places a discount upon injurious groupings and renders their tenure of life precarious. In performing these services, it gives the individual members of valued associations greater liberty and security: it relieves them of hampering conditions which if they had to cope with personally would absorb their energies in mere negative struggle against evils. It enables individual members to count with reasonable certainty upon what others will do, and thus facilitates mutually helpful coöperations. It creates respect for others and for one's self. A measure of the goodness of a state is the degree in which it relieves individuals from the waste of negative struggle and needless conflict and confers upon [them] positive assurance and reënforcement in what [they] undertake.[85]

This practice, of shaping mediatees' relationships with other mediatees so that they promote liberty, equality and collective self-determination, is the practice of *governing*. To govern is to settle on, implement, and enforce the constitutive norms of an institution or community. Algorithmic power is necessary for algorithmically-mediated social relations to be governed in accordance with the values of liberal egalitarian democracy. Of course, algorithmic intermediaries must not bear the *whole* burden of governing power relations between mediatees—law remains crucial, and the role of the state ineliminable. But laws regulating algorithmically-mediated social relations must often be implemented by the intermediaries. Algorithmic intermediaries are uniquely well-placed to shape the distribution of power in the social relations that they mediate. Whether intentionally or otherwise, they will determine, implement, and enforce the constitutive norms of those social relations. If not conscientiously designed to realise egalitarian social relations, to preserve liberty and support collective self-determination, they will reliably reinforce inequality and individual and collective heteronomy, enabling the strong to prey on the weak, the disingenuous on the gullible.

---

[84] Quoted in Brownsword, 2022: 69.
[85] Dewey, 2016 (1926): 112-3.

Lastly, as well as proving essential to enabling morally defensible algorithmically-mediated social relations, algorithmic power offers us a perhaps unprecedented opportunity to reshape our social structures in line with our values. The power to change society through reshaping social relations must undoubtedly be used with humility and caution, but our subjection to heteronomous precipitates of the past is a perennial brake on the realisation of progressive social values. Thus far, algorithmic intermediaries have reshaped society as a side-effect of the pursuit of profit by a few opportunistic individuals. Imagine what could be done if we designed them with liberty, equality, and collective self-determination in mind.

We should conclude, then, that while some kinds of power relations enabled by algorithmic intermediaries should rightly be challenged and eliminated, algorithmic power itself is not inherently objectionable. Indeed, when used to govern algorithmically-mediated relationships, to ensure that the relationships themselves promote liberty, equality, and collective self-determination, and thereby to enable us to bend our broader social structures towards those values, algorithmic power is morally necessary. From here on, then, we will focus on justifying algorithmic intermediaries' *governing power*.

## 3.4. Justifying Algorithmic Governance

One could be forgiven for thinking this story is yet another version of the familiar tale: this exciting technology can do much good! But it also involves risks! Let's build it so we get one without the other! Thus go a million opening paragraphs to grant proposals and policy briefs the world over. But we're not just weighing benefits and harms here; we're talking about the justification of governing power.

To justify governing power, cost–benefit analysis is not enough. Suppose A has governing power over B, C, and D. If A uses that power for sufficiently justified ends, that might protect B–D's negative liberty, but if A is unconstrained by procedural norms that ensure he robustly acts correctly, then B–D still face the risk of unjustified interference, and so are unfree. And if A unilaterally decides what count as the right ends, and B–D disagree, but cannot hold A to account or otherwise influence his decisions, then clearly relational equality and collective self-determination have been thwarted.

Sometimes, unaccountable power might be A's only means to achieve ends that matter enough to override these objections. Think, for example, of emergency powers assumed in the face of some grave and urgent threat. But outside these emergencies (which should be rare), we should aim not just to override these objections with some countervailing value but to resolve them. Indeed, if resolving these objections is possible at a reasonable cost, then the proportionality of overriding them is irrelevant since doing so is unnecessary.[86] All-things-considered justification of power therefore depends on it being exercised not only for the right ends but also in the right way by those with the right to do so. As well as answering the question of *what* power is used to do, we must also ask *how* is it being used? And *by whom*?

In my view (which I cannot defend at length here), *all* power relations are subject to these three distinct questions.[87] However, the *how* and *who* questions are clearly especially stringent when *governing* power is in question.[88] We hold rulers to higher standards of procedural legitimacy and proper authority. Criteria of procedural legitimacy determine how power is exercised, forcing rulers to abide by robust standards. These procedural limits are instrumentally valuable—they should lead to better decisions by the powerful and protect us against both actual unjustified interference and the risk of it. But they also matter non-instrumentally because they reassert relational equality by giving the subjects of power resources with which to hold it accountable. If A has power over B, but A must abide by standards of procedural legitimacy, then B also has power over A, since A's power is limited, and B can force A to comply with those limits. And [A, B, C, D] cannot be collectively self-determining if B–D cannot hold A's decision-making to account.

Even if governing power is being used for the right ends and according to the proper procedures, we can still ask whether those with power have a right to it. This is the question of *authority*.[89] If A has authority over B–D, then A has the right to govern B–D, and they have a duty not to resist her doing so. A's authority also implies (but does not entail) that B–D have pro tanto reasons to comply with A's directives.[90] Some think that if A's power meets the 'what' and 'how' standards, it meets the 'who' standard by default. This is, I think, especially likely to be false when rulers shape power relations between other people. Benevolent, procedurally legitimate authoritarianism remains authoritarian and is inadequate for governing. For example, consider the Facebook Oversight Board—its decisions often seem substantively justified (or at least reasonable) and procedurally impeccable. But the Board's authority derives from a unilateral decision by (then) Facebook to create it. It lacks authority to govern.

In my view, relational equality and collective self-determination together entail that the ultimate source of the authority to exercise power is the community subject to that power. The authority to exercise power must therefore derive from *authorisation* by that community through democratic procedures.[91]

The foregoing gives only a very general framework for the justification of algorithmic intermediary power. But even this limited framework gives us reason to question the prevailing approaches to 'ethical AI' at the forefront of technology development. Companies developing fine-tuned LLMs for Generative AI Systems tout their successes at aligning them with human values, but these approaches at best focus narrowly on improving the models' substantive performance—addressing the 'what' question. This does little to answer the 'how' and 'who' questions—nor is that a coincidence, given that these companies invariably settle questions of safety unilaterally and intransparently, in ways that would not pass even the loosest standards of procedural legitimacy and proper authority.

---

[86] Lazar, 2012.
[87] See Lazar, 2024.
[88] When A's power over B derives from A operating within a certain domain over which A properly has unilateral authority, then the 'who' and 'how' questions are trivially satisfiable—'by A' and 'however she likes'.
[89] Most political philosophers refer to this as legitimacy, not authority. Both terms of art are valid; I think authority is the more linguistically felicitous—legitimacy literally means constrained by rules, which is more obviously connected with procedural legitimacy than with the right to govern.
[90] See Applbaum, 2010.
[91] Compare related views of Viehoff, 2014.

Political philosophy now faces the urgent challenge of going beyond 'aligning' algorithmic intermediaries with our substantive normative aspirations, to considering whether and how they can be procedurally legitimate, and invest governing power only in those with authority to wield it. The task ahead is vast; in the rest of this Lecture, I offer a first pass at the kinds of questions before us.

## 4. AUTHORITY, COERCION, AND PRE-EMPTION

Because algorithmic governance is interestingly different from other, more-established means of governance—sometimes just due to speed and scale, sometimes due to its interesting new combination of distinct modalities of power—we can expect a theory of justified algorithmic governance to provide fertile ground for philosophical inquiry. Analytic political philosophy's pre-eminent theories of justified governing power are tailored to specific means and modalities of governance: the law, in particular in its extrinsic modality, as implemented by the state. Will those theories extend naturally to cover algorithmic governance, which is more intermediary than extrinsic, and is typically exercised by private entities, not nation-states? If not, what deeper moral foundations must we excavate, in order to ground both extant political theories and those adapted to algorithmic governance? And can we determine, in general, whether algorithmic governance is presumptively easier or harder to justify than the means and modalities of governance that analytical political philosophy has typically centred?

The justification of authority affords a compelling beginning. In both contemporary political philosophy and its early modern roots, the question of authority fundamentally concerns the state's right to exercise extrinsic power over its subjects, through making laws with which those subjects are bound to comply. Some focus on whether subjects have an obligation to obey the state's laws.[92] Rawls' liberal legitimacy principle asserts that the state may coerce its citizens only in accordance with a constitution, the essentials of which they cannot reasonably reject.[93] Raz defends the authority of law, arguing that we have content-independent reasons to obey it insofar as doing so reliably improves on independently judging what we have most reason to do.[94] Elizabeth Anderson looks beyond the state, considering employers' authority to govern their employees—but she also focuses on the ability to issue coercive directives backed by threats of force, arguing that since firms, as well as states, have this ability, the same normative standards should apply.[95] Beyond normative political philosophy,

---

[92] Simmons, 1979.
[93] Rawls, 1993.
[94] Raz, 1979, 1986.
[95] Anderson, 2017.

Weber also argued that state's defining feature is its claim of rational-legal authority for its use of violence against the people within a given territory.[96] Each of these theorists interpret the basic question of who has the right to exercise power as centrally concerning the exercise of extrinsic, coercive power through the law.[97]

Algorithmic intermediaries also coerce, and distinguishing between extrinsic and intermediary coercive power can be challenging—the line between them is blurry. But they enable a different kind of power too, which shapes our behaviour not by issuing directives and altering incentives, but by determining what is and is not possible within the social relations they constitute. This 'pre-emptive governance' has three dimensions.[98] First, what Roger Brownsword calls 'technological management': the choice simply not to design non-compliant options, or to design them away. Meta's personal space boundary in its Horizons virtual world is the paradigmatic case here. Second is the provision of compliant options, described above as *productive* power. This productive power is central to the law as well as to algorithmic governance (for example, law creates certain kinds of socially-recognised relationships, like marriages). Third is what Karen Yeung has called 'hypernudging', in particular, affording preferred options and frustrating dispreferred ones in a manner that dynamically updates and responds to user histories and actions.[99] Hypernudging too is named for its counterpart in our non-algorithmic lives but can be personalised and perfected in the Algorithmic City.[100]

Pre-emptive governance is not unique to the Algorithmic City—'regulation by design' is a familiar practice, as those in power use locks, barriers, productive power, and choice architecture to shape people's behaviour.[101] And yet there are important differences between these technologies of power. Physical and legal infrastructures are costly to develop, hard to adapt, and often have unexpected affordances—we cannot anticipate how they will be used or easily adapt them when they have unintended consequences. Walls must be built and cannot easily be moved. Laws must be introduced, promulgated and enforced—and depend on a measure of publicity, stability, and voluntary compliance that militates against repeated revision. Physical artefacts and laws are also limited in the control they can exercise: walls can be surmounted or dug under; constitutive laws can determine whether a given kind of social relation is legally recognised but cannot (without costly enforcement) prevent people from participating in the legally-unrecognised counterpart. By contrast, algorithmic intermediaries operate in an environment that is, in principle, perfectly malleable and dynamically updatable, which can be fine-tuned and personalised at a low cost, and which can exercise in-principle perfect control of behaviour.

---

[96] Weber, 2019.

[97] While many twentieth century political philosophers recognised the constitutive role of law (see e.g. Rawls, 1955; Hart, 1961), I am aware of no account of the justification of authority that asks what gives the state the right to play that constitutive role (as distinct from its right to coerce us, or do other things that it would be impermissible for us to do to one another).

[98] On pre-emption, see among others Brownsword and Yeung, 2008a; Yeung, 2008; Zittrain, 2008; Nissenbaum, 2011; Hildebrandt, 2015.

[99] On technological management, which Lessig called 'coding problems away' see Lessig, 2006: 15; Brownsword, 2022, and in general Brownsword and Yeung, 2008b. On productive power, see Bucher, 2018, and on hypernudging see Yeung, 2017.

[100] Arguably hypernudging is more of a departure from offline nudging than Yeung allows, given that the focus of nudging, in the terms introduced by Sunstein and Thaler, was emphatically to influence people's choices paternalistically without manipulating them, whereas hypernudging can be seen as straightforwardly manipulative, and is extractive rather than personalistic. Thanks to Kyle van Oosterum for helping me see this.

[101] Yeung, 2008.

Of course, in the Algorithmic City as it is now, pre-emptive governance is imperfect. Sunk costs, such as legacy code, cannot without massive expense be repurposed. No software system is ever perfectly secure against attacks.[102] Successful governance, therefore, still depends heavily on algorithmic and legal coercive power. But the advent of the Algorithmic City invites a thought experiment in which these practical limitations are overcome. So consider now an imaginary counterpart of the Algorithmic City, which we can call Pre-Emptopolis.[103]

Pre-Emptopolis is constituted by a unified set of algorithmic intermediaries. Instead of being an occasional retreat from our offline lives, most of its denizens spend most of their time there—imagine the most ambitious projections of what the 'metaverse' might be like. In Pre-Emptopolis, our options are pre-emptively shaped to ensure they comply with some background set of norms. We therefore lack the choice not to comply. For example, we cannot assault others when they aggravate us—the attempt to do so is impossible to execute. We cannot steal others' virtual property. Our communications are monitored by highly advanced language- and image-processing models that recognise and prevent communications that do not comply with the designers' chosen communicative norms. Algorithmic intermediaries perfectly constrain our behaviour, so we have no alternative but to behave in the ways selected by Pre-Emptopolis' designers.

If political philosophers were correct that authority concerns only the right to coerce, the right to make law, the duty to obey the law, or the authority of law, then questions of authority would not arise in Pre-Emptopolis, because it has no role for coercion, law, or indeed obedience. *The duty to obey the law presupposes one has the option to disobey.* This is why the duty matters: law depends on us deciding to comply, even if we could do otherwise and get away with it. But in Pre-Emptopolis, we do not need law; non-compliance is impossible, and conformity can be ensured without voluntary compliance. And yet Pre-Emptopolis very clearly raises deep and concerning questions of authority. What gives its designers the right to decide which actions and communications to admit and which to 'code away', which to encourage and which frustrate? If Pre-Emptopolis, like present-day algorithmic intermediaries, were run by a private, for-profit company, their authority to exercise this kind of power would be deeply morally suspect.

---

[102] Grimmelmann, 2004.

[103] Thanks to Sean Donahue for suggesting this name. Brownsword, 2015 develops a related hypothetical: a whole society (not just algorithmic intermediaries) governed by means of technological management alone. In Brownsword, 2022, his synthesising monograph on this topic, he starts with jurisprudence, not political philosophy. He considers standards of legality, the rule of law, and legitimacy, which roughly correspond to what I call procedural legitimacy, authority, and substantive justification. Legality is interpreted as a conceptual or pragmatic implication of the idea of governance by rules. The rule of law is understood through the lens of existing legal institutions: X governs in accordance with the rule of law if the law of the land licenses X to govern. Legitimacy is about consonance with universal, constitutional, or democratically-chosen substantive values. In each case, Brownsword's approach is positivist where mine is fundamental: procedural legitimacy is not, in my view, just a conceptual implication of governance by rules, but is directly grounded in the values of liberty, equality, and collective self-determination; authority *can* be derived by authorisation from institutions that are themselves appropriately authorised, but our task is to determine the basis of authority, not to assume it; and I directly argue for the substantive values algorithmic governance should aim at, rather than argue indirectly that it should support a community's values, whatever they are. In general, I think that Brownsword's work, while brilliantly surfacing the dimensions of algorithmic governance, ultimately takes the law as fundamental and seeks to understand algorithmic governance in those terms, rather than seeing governance as fundamental and the law as just one modality of governance among others. I highlight further differences between our views in footnotes below.

Pre-Emptopolis is a thought experiment. Though some aspire to design an internet of the future in which 'can't be evil' trumps 'don't be evil', experience suggests that this goal is likely to be unachievable with existing technologies or their likely extensions.[104] I aim not to advocate for or against Pre-Emptopolis. The thought experiment shows that the question of authority cannot be reduced to justification of the right to coerce or impose laws or the duties correlative to those rights. This has two important implications.

First, our existing theories of justified authority are incomplete. States also exercise pre-emptive power. Do the established accounts of justified authority adequately ground that practice, as well as the extrinsic coercive power on which they have focused?

Second, and more generally, the argument shows that theories of justified authority must deploy a broader, more overarching concept to encompass both pre-emptive and coercive power. I think we should focus on the right to govern.

To govern is to settle on, implement, and, if necessary, enforce the constitutive norms of an institution or community.[105] An institution is (for our purposes) a social object composed of a set of roles occupied by people who stand in social relations with one another, where the institution's norms determine the content of those roles and how they interrelate. I mean 'norms' in the most generic sense, encapsulating any normative propositions. Governance has three general stages: deciding the norms that shape how the institution operates, putting them into practice, and enforcing them ex post. This could be through imposition of sanctions or through finetuning the pre-emptive design of the system to eliminate non-compliance that slips through the cracks.

Rulers can adopt different means and modalities of governance. In human history, extrinsic coercive legal power has typically been central. We can imagine a society without pre-emptive governance, in which all norms are enforced coercively. Conversely, we can imagine Pre-Emptopolis, in which coercive governance is unnecessary because all non-compliant options have been designed away. Both approaches to governance raise the same central questions of authority: what gives you the right to settle on, implement, and (if necessary) enforce these norms? Indeed, pre-emptive governance arguably raises these questions more acutely, since the logic of pre-emption invites governing power to obscure itself and fade into the background where it cannot easily be challenged.[106]

---

[104] For some scepticism about the prospects of perfect regulation through code, see again Grimmelmann, 2004.

[105] Some scholars would describe this as 'regulation', not 'governance'. Governance seems the more general term; it is more useful to reserve regulation to refer specifically to the modality of governance that involves a duly constituted body establishing rules, then monitoring and enforcing compliance with those rules. See Black, 2002; Yeung, 2018; Brownsword, 2022: 26.

Governance involves deciding the constitutive norms of an institution, shaping power relations between those subject to those norms, adjudicating between them, and enforcing compliance, either pre-emptively or coercively. Each stage involves exerting power and making substantive practical and moral judgements. To govern is to assert one's right to decide right and wrong and to settle empirical disputes.

Given the persistence and pervasiveness of deep moral (and epistemic) disagreement, unilaterally assuming the authority to settle these disputes affronts relational equality: it disrespects the equal moral status of the people being judged, and their in-principle equal ability to make these kinds of judgements.[107] It may be easier to reach an objectively valid answer in purely empirical disputes. But rulers often have to be 'arbiters of truth'.[108] Unilaterally adopting that role creates relational hierarchy between the arbiters and the arbitrated. This can be mitigated or dissolved through proper authority and procedural legitimacy, which ground the arbiter's authority to make those determinations and provides boundaries that channel and limit their power so that the arbitrated can hold them accountable.

Governing power directly implicates collective self-determination too. We cannot collectively shape the shared terms of our social existence if algorithmic intermediaries unilaterally decide them for us. The constitutive norms of the Algorithmic City are fundamental elements of our shared social lives. If decided for us, not by us, we are not collectively self-determining. Likewise for monitoring and enforcing those norms—if we are subjects of that enforcement rather than its ultimate authors, then we are not collectively self-determining.[109] This is nowhere more apparent than in the early days of a new technology being deployed, as people detect how they are being governed without having any say in the matter— as an example, consider the furore over content moderation, or how ChatGPT skews progressive and rejects invitations to endorse discriminatory judgments.[110]

Pervasive moral disagreement suggests a further argument that governing power should meet standards of authority, indirectly grounded in negative liberty. Adjudicating right and wrong involves the risk of making a wrong decision and interfering unjustifiably in someone's life. What could justify that risk? One crucial factor is whether you act on behalf of the community you govern. If you do, you can plausibly appeal to two kinds of reason that apply to them to justify the risks you are taking. First, if your authority to govern is grounded in democratic authorisation, then those subject to it have implicitly authorised your power over them, diminishing their complaint against you (assuming you are procedurally diligent in your exercise of governing power).[111] When those with proper authority govern us, we, to some degree, indemnify them against their (procedurally legitimate) substantive mistakes. Second, if your authority is grounded in authorisation by the broader community, then you are not acting as a private citizen, motivated by justice, but on behalf of the whole community. Reasons that apply to them—to see that their community is responsibly governed—count in favour of your diligent pursuit of that duty and can weigh against the risks of unavoidable error.[112]

---

[106] Brownsword (2022: 199) thinks that in a scenario like Pre-Emptopolis the concept of legal authority would be redundant—this is correct, but ignores the fact that the concept of authority itself is not exclusively about the authority of the law. Even if Pre-Emptopolis were governed *only* by algorithms, with no (living) designers to hold accountable, we must still ask whether these algorithmic systems have the right to govern, and if we have correlative duties not to resist or challenge their authority.

[107] Rawls, 1993.

[108] Klonick, 2017; Douek, 2021.

[109] Rousseau, 1762.

[110] Rozado, 2023.

Political philosophy has mostly understood authority as the right to exercise coercive power to make and impose laws. Although states and other powerful entities have always also used pre-emptive power to shape behaviour, this has been largely overlooked in the justification of authority. The Algorithmic City invites us to consider pre-emptive governance more closely, recognising it as a modality of governance on par with coercion. It leads us to excavate the value of authority and find the deeper underlying principle covering pre-emptive and coercive governance. If political philosophy has a central question, it should be: what gives some the right to govern others? And because the need for correlative obedience is a contingent feature of one modality of governance, which can in principle be eliminated by fully algorithmic governance, we have conclusive evidence that the duty to obey the law is not the correlate of the right to govern (as Simmons thought it was).[113] The right to govern is, instead, correlated with a duty not to resist or interfere with that governing power.[114]

## 5. PROCEDURAL LEGITIMACY AND ALGORITHMIC GOVERNANCE

Familiar accounts of procedural legitimacy are inextricable from the prevailing modalities of governance in the historical state. They are legalistic and are mainly explored by legal rather than political philosophers.[115] How do these norms apply to pre-emptive algorithmic governance, which needs neither to specify nor to enforce rules? How do they apply to algorithmic coercion, with its promise of perfecting surveillance and enforcement? The first step is to excavate the values beneath procedural legitimacy, to see whether and how they apply to algorithmic governance that differs (whether in degree or kind) from the legalistic governance for which current procedural legitimacy norms were tailored.[116]

---

[111] This is also true for those who had the opportunity to participate in the democratic process, but chose not to take it. Thanks to Garrett Cullity for discussion here.
[112] Lazar, 2016.
[113] Simmons, 1999.
[114] Applbaum, 2010; Applbaum, 2019. Note, though, that there is a difference between resisting a governing power and resisting a particular decision or rule imposed by that power. The latter isn't at all implied by the right to govern.
[115] Contrast my approach here with that of Brownsword, 2022: 58ff. Brownsword again takes governance by laws as fundamental, and explores how the affordances of technological governance enable or limit fulfilment of principles of legality. I think the values underpinning procedural legitimacy are fundamental, and principles of legality are just one possible expression of those values for one technology of governance.
[116] Cohen, 2003.

At root, procedural legitimacy presupposes that the governed should be able to limit their rulers' power, holding them to strict rules, ensuring we can verify compliance with those rules and challenge them if they overstep. This protects us against the risk of wrongful interference and reasserts equality between our rulers and us. They govern us, so they have power over us, but they must abide by these limitations, so we have power over them.

When our rulers use the law to govern, we constrain them by imposing ex ante, in medias res, and ex post constraints on their exercise of power. Laws must be public, subject to notice and comment before they are applied; they must be applied consistently and fairly; those subject to the law must be able to contest adverse decisions and hold accountable those charged with applying the law. Algorithmic intermediaries do not govern us through law. But they must still abide by criteria of procedural legitimacy. So, rather than taking these legalistic standards as basic, we must instead dig beneath them to identify the underlying values towards which they seek to direct legal procedure, so we can use those values to guide procedural standards for algorithmic governance.

I suggest that the key underlying constraints on governing power are authorship, transparency, consistency, accountability, and resistibility. Authorship is expressed in the requirement that laws be made public ex ante so that we know they apply to us and we have the opportunity to influence them. Transparency runs throughout the three stages of procedural legitimacy; it is essential to understand how these laws are applied and by whom. The demands that laws should be implemented consistently and those who implement them should be accountable for misconduct clearly appeal to values that run deeper than the forms of particular legal procedures. Quixotic and unaccountable governance is objectionable, whatever its form.[117]

The preceding values can all be easily read off from familiar approaches to procedural legitimacy and the rule of law. A further value is easy to ignore but becomes especially important when considering algorithmic governance. Formal rules and procedures are not always sufficient to hold the powerful in check. Robust limits on governing power require that its subjects have reasonable opportunities to *resist* that power when other means for constraining it prove unsuccessful.[118] Resistance can be targeted either at the ruler or at some specific set of actions that they have taken. Note that only the former involves questioning the ruler's authority.

With these underlying values in mind, we can see that legal governance has two 'accidental virtues'. An accidental virtue arises when a practice or technology's effective functioning gives us a moral benefit for free, often due to some practical and contingent limitation of that technology. For example, our inability to predict who will be lucky and who unlucky has historically meant that insurance has the accidental virtue of realising a measure of fairness between those groups: for insurance to work for anyone, the lucky have to subsidise the unlucky. If we then become able to predict who is lucky, and who unlucky (for example through genetic sequencing), then insurance can function for the lucky without them subsidising the unlucky. The accidental virtue is lost, and we have to intentionally engineer it—at a cost.

---

[117] Nissenbaum, 1996; Binns, 2018.
[118] Resistibility obviously matters most in our actual non-ideal circumstances, but would I think continue to be important even if we had an ideal democratic government, with every provision made for formal political participation. It provides an extra-institutional backstop, which protects against any capture of existing institutions. Thanks to Massimo Renzo for discussion here.

Law's accidental virtues include publicity and resistibility. Law aims to secure compliance without resorting to the costly expedient of oppression. It succeeds when we internalise those norms and comply with them, even when we reasonably believe we could get away with non-compliance. For law to achieve those goals in our analogue lives, it must be public.[119] People cannot internalise law's injunctions if they do not know what it says, and their incentives will change if they do not know how the law is being applied. But if the laws are public, we have at least *some* opportunity to call attention to them and criticise them. This is not sufficient for authorship, but it is necessary. And as transparency in law's application is essential for it to calibrate our voluntary behaviour, it also enables us to determine whether the law is being applied consistently, and affords a necessary condition for ensuring accountability for abuses of legal power.

Both due to its public nature and because the nature of law presupposes the possibility of non-compliance with the law, the law also affords resistibility. Not only is it possible to resist the law, but one can do so publicly in ways that invite others to support or join in one's cause. We can choose public disobedience with specific laws or rejecting the authority of the entity making law. The public spaces of the analogue city themselves afford public resistance, enabling us to come together with other disobedients and demand change.

I have argued elsewhere (as have many others), that algorithmic governance lacks these virtues—especially publicity.[120] Algorithmic governance often relies on secret, complex, and intrinsically inscrutable computational systems that even experts in computational systems cannot understand. This makes publicity a forlorn hope in many cases. But as well as showing that systems based on ML are intrinsically opaque, we can also show that both pre-emptive and coercive algorithmic governance lack the accidental virtues of publicity and resistibility, because they can both be equally effective without them. I'll take them in turn.

In Pre-Emptopolis, transparency is unnecessary because we can secure compliance without people knowing which rules apply to them or how they are applied just by designing non-compliance away. Since we need not publicly announce rules to ensure compliance, Pre-Emptopolis can be successfully governed without its denizens exercising any authorship over their options. Without enforcement and transparency, consistency and accountability seem to lack actionable referents. Indeed, governance in Pre-Emptopolis can be algorithmically personalised to each individual, so it is unclear what consistency would even mean. And, perhaps most importantly, our opportunities for resistance in Pre-Emptopolis are substantially limited by what is allowed us by its rulers.[121]

Law has to be public to be effective; pre-emptive governance can be wholly invisible.[122] Disobeying the law is always an option (however costly). This is no longer true in Pre-Emptopolis, where our choices are pre-determined. And nothing about Pre-Emptopolis necessitates the provision of public space for forging coalitions to challenge the authority of Pre-Emptopolis' rulers. The denizens of Pre-Emptopolis are presumptively passive subjects whose ability to resist their overlords within that platform is conditional on the latter's provision of the option of resistance. Indeed, even if public 'spaces' are provided, they still afford less scope for resistance than the analogue city; algorithmic intermediaries enable some kinds of coalition-building but otherwise fragment us and undermine collective action.[123] For example, part of building a coalition is signalling mutual commitment to other prospective members. Being present in person—at personal risk—is a costly signal of commitment and facilitates trust. Merely digital interaction is a cheap signal, and trusting those on the other side of an algorithmically mediated relationship is much harder.

---

[119] Indeed, some legal philosophers think that meeting criteria of procedural legitimacy is a necessary condition for something actually *being* law (Fuller, 1964).
[120] Lazar, 2024.
[121] Brownsword (2022: 157) draws attention in passing to the removal of conscientious objection as an option (I think civil disobedience is the more salient category).

Setting aside the hypothetical case of Pre-Emptopolis, consider the now very real case of LLMs, trained to govern their users by means of reinforcement learning with computational feedback. Since the law aims to shape people's behaviour, it must not only be public, it must also be interpretable, predictable, and action-guiding. We need to know what we need to do in order to comply with it. Algorithmic intermediaries based on safety-tuned LLMs don't need to proceed by way of our will. And they could not if they sought to, because of the peculiar way in which the underlying models are trained to adopt and implement the values by which they govern. They may herald the advent of a new modality of governance. Legal governance proceeds by way of rules on the one hand, and standards on the other, where rules precisely specify permissible behaviours, while standards specify desirable attributes subject to the discretionary evaluation of the governing party. This discretion makes governance by standards presumptively harder to render procedurally legitimate than governance by rules. But we mitigate that disadvantage by ensuring that those who exercise discretion are properly authorised to do so, are reliable and trustworthy, and can be individually held to account if they abuse the freedom that this discretion gives them. LLMs govern by way of often secret rules—for example, when specific prompts are filtered based on keyword-matching. And they govern by standards, learned through the process described above. The principles that provide these standards are almost silly by human lights (Anthropic's 'Constitution' for its Claude model includes Apples Terms of Service alongside principles drawn from the Universal Declaration on Human Rights).[124] They are incredibly ambiguous. And the model is left to determine how to weigh them against one another by its own lights. We have no grounds to trust these models—they do not perceive the world in the way that we do, and they have no other attributes that warrant trust, unlike the humans in whom we invest discretionary authority. And they obviously cannot be held to account for bad performance in their exercise of discretion. 'Governance by prompt' is a new and weird modality of governance, which is presumptively at odds with our expectations of procedural legitimacy.

---

[122] 'We should worry about this. We should worry about a regime that makes invisible regulation easier; we should worry about a regime that makes it easier to regulate.' (Lessig, 2006: 151). See also Gillespie, 2018: 179.

[123] Tufekci, 2017; Vaidhyanathan, 2018; Draper and Turow, 2019.

[124] https://www.anthropic.com/index/claudes-constitution.

Algorithmic governance need not only be pre-emptive. In fact, algorithmic intermediaries enable the in-principle perfection of coercive enforcement, which becomes much easier if you control every aspect of the environment in which potentially non-compliant behaviour (which has not been pre-empted) takes place.[125] This was Jonathan Zittrain's warning about the transition in digital culture from the open internet and general-purpose computing to appliances managed by private companies.[126] It has even greater force in our world of centralised platforms and algorithmic intermediaries, as they control our access to each other.[127] It would be more potent still in visions for algorithmic governance of the metaverse and web3, and if we become reliant on Language Model Agents to serve as universal intermediaries to digital technologies. Advances in AI over the last decade have only enhanced the promise of pairing perfect surveillance with comprehensive enforcement.[128]

Technological limitations notwithstanding, algorithmic governance still enables more comprehensive enforcement than legal governance. This means the scope for resistibility is radically limited. Of course, some hackers and jailbreakers may always be able to evade these governance measures.[129] But while this might help them secure private individual benefits, or participate in a peripheral subculture, the value of the Algorithmic City lies in how it mediates social relations at scale, and algorithmic intermediaries' strength is in coordinating collective action, corralling audiences and markets. The possibility of interstitial non-participation by the technologically savvy does not affect the degree to which those who want to participate in the same shared space are subject to perfectible enforcement.

Next, crucially, coercive algorithmic governance like its pre-emptive counterpart does not rely on public promulgation and widespread voluntary compliance to be effective. Because behaviour can be comprehensively monitored at relatively low cost, and interventions and penalties can be automated, algorithmic governance can enforce compliance without either widespread internalisation of the law or costly public shows of force. Instead, it can proceed through the invisible imposition of conformity. Publicity is therefore not called for (which further undermines resistibility).

Both pre-emptive governance and (the prospect of) perfect coercive enforcement in the Algorithmic City induce the hallucination that we can settle on the right norms, design the ideal algorithmic system to implement them, and then forget about governance. This hallucination is especially prevalent among those focused on aligning LLMs, or among the advocates of web3, who persist in the belief that sufficient care to mechanism design can ultimately do away with politics. But it is emphatically wrong. Contestation and resistance, not just of particular judgements but of the norms they apply, are essential both for moral progress and for respecting our standing as free and equal members of the moral community. And denying people the option of resistance potentially undermines the justification of authority, since having but not taking that option indicates at least a minimal level of endorsement of governing power. The advent of algorithmic intermediary power reveals how our analogue governance practices have the accidental virtue of affording this resistance. Algorithmic governance facilitates denying us the shared experience that undergirds collective action, eliminating options to come together and protest, monitoring our attempts to form a common mind, and applying penalties to those who do not comply. Though our theories of procedural legitimacy have been coloured by their focus on law, their essential commitment is to limitations on governing power. Algorithmic governance invites our rulers to obviate all limits. This does not mean the Algorithmic City inevitably entails authoritarianism. Affordances are not guarantees. But *if we do not intentionally counteract them*—or indeed seek to exploit them for private profit—then authoritarianism is correspondingly more likely.

---

[125] Lessig, 2006; Zittrain, 2008; Kerr, 2009.
[126] Zittrain, 2008.
[127] Giblin and Doctorow, 2022.
[128] For some scepticism on the prospects for algorithmic enforcement of norms including copyright and content moderation, see Suzor, 2019; Gillespie, 2020; Gorwa et al., 2020; Keller, 2021.
[129] Grimmelmann, 2004.

## 6. JUSTIFICATORY NEUTRALITY

Algorithmic intermediaries exercise governing power across every domain of our lives, inevitably raising innumerable questions of substantive justification: in each case, we have to ask just what algorithmic intermediary power should aim at. In Lecture II I broach this question for the digital public sphere. In this section, I focus on one overarching question of substantive justification: whether and how we can aspire to justificatory neutrality in the Algorithmic City. The central idea is simple. Legalistic governance affords justificatory neutrality; algorithmic governance makes it at least somewhat harder to achieve.

Governing authorities adjudicate right and wrong under reasonable moral disagreement. The moral risks of doing so can be mitigated by satisfying principles of proper authority and procedural legitimacy. But one of liberal political philosophy's central tenets is that this is not enough. We must also limit scenarios where the state's judgements of right and wrong involve weighing in on the side of one or another competing conception of the good. The state should strive for neutrality among those who reasonably disagree about morality and other such central questions.[130] While neutrality of *impact* (or 'outcome neutrality') is unattainable, neutrality of justification remains a widely-shared goal, grounding governing decisions on appeal to things that we can broadly agree upon rather than just imposing the values of a majority. Even if this goal may prove unattainable,[131] we can still aspire to it.

Extrinsic governance affords justificatory neutrality. The basic paradigm is to secure sufficient agreement on the parameters for behaviour within some domain, then leave people free to decide how to behave within those parameters. We provide physical and institutional spaces for unmediated interaction and then trust that these extrinsic constraints—Dewey's river banks—will be enough to channel us towards broadly acceptable social outcomes without having to directly invest our choices with the rulers' values, and while creating a broadly level playing field for different ideas and values to be advanced and adopted.

Justificatory neutrality seems feasible for extrinsic governance for two reasons:

---

[130] E.g. Rawls, 1999; Quong, 2011; Patten, 2012.
[131] Sher, 1997; Arneson, 2003.

first, we have a greater chance of reaching broad agreement on the extrinsic boundaries intended to rule out the worst kind of conduct than if we had to, for example, rank or otherwise weigh in on all sorts of permissible conduct. Second, these minimal parameters can be *silent* on specific substantive questions, taking no view. Of course, if the law (say) is silent on X, then by default, the law permits X.[132] But this kind of implicit permission is distinct from the active endorsement of X—for example, by funding it or protecting it from interference—and its prohibition. It still meets the goal of justificatory neutrality, if not outcome neutrality.

The essence of governance is deciding on (and implementing) the constitutive norms of an institution or community. Extrinsic governance allows the ruler to avoid deciding on a wide range of constitutive norms. By contrast, intermediary governance constitutes the social relations that it governs. As a result, it enables, perhaps even necessitates, a kind of totalising governance, which prevents it from meaningfully prescinding from judgement concerning *any* substantive normative questions.[133] Every option has to be designed.[134] This makes justificatory neutrality much harder to achieve in three ways.

First, the designers of algorithmic intermediaries have *so many more* decisions to make than those who govern social relations in the analogue city. As well as having to decide, for every option, whether to endorse or prohibit it (lacking the intermediate option), they must also determine which kinds of act to afford, which frustrate, which to promote and which demote. Of course, the analogue city has its affordances too, but, as argued above, they are much harder to monitor and fine-tune than those of the Algorithmic City.

Consider the contrast between the governance of online speech and the pre-internet 'marketplace of ideas'. The latter allowed the state to provide starting-gate conditions and get out of the way. But (as I argue in Lecture II) algorithmic intermediaries shaping the digital public sphere must decide not only what kinds of speech are admissible into the public sphere, but also how that speech will be ranked, and how visible it will be. They cannot escape this decision: whatever they do, they will promote some communications at the expense of others; if they do not do so conscientiously, we know very well that the most extreme and emotive content will rise to the top.

Or consider algorithmic 'markets', which are radically different from markets in the analogue city. The same underlying principles of mechanism design might theoretically apply, but in the algorithmic setting, the degrees of freedom are much greater because the designer can shape so much more: both interactions between the two sides of the market and the degree to which information is shared with the mechanism designer.¹³⁵ Algorithmic markets enable routinised arbitrage between those being mediated and extraction of information salient for competition. They involve deciding how to rank both sides of the two-sided market. Of course, analogue markets also encode the values of the societies that deploy them—they are hardly exemplars of justificatory neutrality. But they at least allow establishing a set of starting-gate conditions, which can, in principle, be agreed on antecedently, enabling people to engage in mutually-beneficial exchange without relying on intermediaries. In algorithmic quasi-markets everything is up for grabs; everything is an object of design, so must be an object of normative choice.

---

[132] One of the more annoying aspects of Elon Musk's purchase and destruction of Twitter is that the letter, X, beloved of analytic philosophers, now has a double meaning. I reject Musk's appropriation of this common resource, and as such whenever I use 'X', I intend it to play its role as a variable in analytic philosophy. I will refer to Musk's platform as X (formerly known as Twitter).

[133] 'We can build, or architect, or code cyberspace to protect values that we believe are fundamental. Or we can build, or architect, or code cyberspace to allow those values to disappear. There is no middle ground. There is no choice that does not include some kind of building. Code is never found; it is only ever made, and only ever made by us.' Lessig, 2006: 6.

[134] Gillespie, 2018: 179.

Algorithmic intermediaries can enable end-to-end encryption, preventing them from recording more than just the metadata associated with algorithmically mediated exchanges (or even obscuring that). But encryption is not a neutral default; it is an active choice. And even when communications are encrypted, they are still architected by the algorithmic intermediary. They can still be subjected to algorithmic governance, as with the use of client-side scanning to prevent sharing of child sexual abuse material, which could be extended to, for example, identify when people are privately sharing misinformation or to disrupt plans to brigade people on social media.¹³⁶

The truth that technology is not neutral has surely reached even the deafest ears.¹³⁷ But its corollary for political philosophy has not been adequately appreciated. Algorithmic governance cannot easily be neutral because it is implicated in what it mediates and because there is no 'natural' baseline to fall back on. So it requires making explicit judgements of right and wrong *everywhere*.

Second, in the Algorithmic City there is no relatively neutral middle ground between prohibiting an option and endorsing it. For an algorithmic intermediary to permit an option X, it has to make it possible for users to X, and thereby becomes at least somewhat implicated in their X-ing. If A knowingly provides B with the means to X, then A must be read as endorsing X, at least to a greater degree than if A merely fails to prevent B from X-ing. The only way not to endorse some option in the Algorithmic City is not to enable it, which is expressively close to prohibiting it, and as such a more forceful judgement than mere permission without endorsement; one cannot say that X is prohibited but that this does not reflect a judgement that one ought not X.

This observation is a corollary of the different defaults that apply to governing the Algorithmic City and its non-algorithmic counterpart. In the analogue city, for any given behaviour X, the default is that X is permitted unless it is explicitly prohibited. Governance operates against a baseline of freedom of action. This is dictated by the reality of our physical lives: our ability to perform a wide range of the most fundamental actions is not conditioned on the law making those actions possible.¹³⁸ In the Algorithmic City, the default is that you are *not able to* X unless it is explicitly enabled. This is dictated by the nature of algorithmic intermediaries—action within them *is* conditioned on the intermediary making it possible. Any given behaviour X can be permitted only if the designers of the relevant algorithmic intermediaries first enable it. So governance operates against a baseline of unfreedom. Of course, in practice, X will often be enabled accidentally, but once X is a known use of the intermediary, the decision to continue to enable it when it could be disabled amounts to intentionally enabling it. Moreover, this is an active form of enabling: algorithmic intermediaries make X possible, they contribute to it. The difference between law and algorithmic governance, then, is that between allowing behaviours to happen, and contributing to them. If A enables B to X, providing the means with which to X, other things equal that implicates A in B's X-ing to a greater degree than if A merely allows B to X.[139]

---

¹³⁵ Birch, 2020; Viljoen *et al.*, 2021.
¹³⁶ Indeed encrypted communications platforms like WhatsApp *are* still subject to algorithmic governance: see https://www.propublica.org/article/how-facebook-undermines-privacy-protections-for-its-2-billion-whatsapp-users.
¹³⁷ Hare, 2022.
¹³⁸ Even though *some* kinds of actions are constituted by law, most are not. The contrast between the Algorithmic City and the analogue city is one of degree here, not of kind—but it's a difference in degree that makes justificatory neutrality harder to achieve.

That extrinsic governance operates against a baseline of freedom provides one explanation for why justificatory neutrality is an appealing alternative to outcome neutrality. If our rulers implement a neutrally-justified extrinsic governance regime which has non-neutral impacts, then they can defensibly appeal to people's intervening voluntary agency to explain those impacts and effectively deny that their regime is responsible. If a neutral governance regime leads to non-neutral outcomes because of the free and voluntary choices of those subject to it, then those non-neutral outcomes are acceptable at least to that extent.

This argument is unavailable for intermediary governance in the Algorithmic City. Its rulers cannot to any significant degree disown or distance themselves from outcomes realised by people's voluntary choices because they chose among options that their rulers selected. Intermediary power constitutes the relationships that it governs. Designers of algorithmic intermediaries must take responsibility for what people do within those relationships by way of those intermediaries, because they have provided them with the materials with which to act. It is therefore much harder to separate justificatory neutrality from outcome non-neutrality in the Algorithmic City than in the analogue city. The appeal to justificatory neutrality seems like a sham if your design decisions precisely enable and constitute differential impacts on different conceptions of the good. This is vividly clear in the debates over how Generative AI systems should work. While some open source developers and CEOs have tried to argue that all responsibility for the outputs of text and image-generation models lies with the users, the leading AI firms all recognise their deep responsibility to mitigate these risks, and to prevent their tools from being used in harmful ways. Indeed, the principal success of ChatGPT was precisely that it was fine-tuned to prevent toxic and otherwise harmful generations, and Anthropic's signature innovation is the development of Constitutional AI, described above. These are paradigmatic examples of intermediary governance; they are rightly prioritised by the companies because their ability to govern user behaviour by training the model, as well as the way in which the model is implicated in the communicative acts that it mediates, mean they cannot escape responsibility for what people do with their models.

---

[139] My argument presupposes a broadly deontological approach to the responsibilities of governance, drawing a distinction between doing and allowing—it matters, in my view, whether and how algorithmic intermediaries contribute to wrongful behaviour. For consequentialists, all that matters is whether that behaviour is preventable, not how it is caused. Consequentialists therefore cannot meaningfully distinguish between justificatory and outcome neutrality. I think this tells against consequentialism as a foundation for political philosophy, but cannot defend that point here. Many thanks to James Grimmelmann and Massimo Renzo for discussion on this point. On doing and allowing, see Quinn, 1989.

Third, another foundation for the appeal of justificatory neutrality in the analogue city is that, in at least some domains, it reliably leads to more-or-less positive social outcomes. Rawls favoured political liberalism because he thought it necessary to realise social stability (for the right reasons) in the context of reasonable pluralism. But when we try to create something like this permissive environment in the Algorithmic City, it invariably results in a toxic mess. There are many reasons for this, most concerning the difficulty of sustaining robust interpersonal norms in a virtual environment characterised by anonymity, fleeting encounters, and platforms that reward confrontation. Whatever the explanation, laissez-faire in the Algorithmic City invariably amounts to laissez suffer—and algorithmic intermediaries cannot escape responsibility for these toxic outcomes by appealing to the intervening agency of their users because they have knowingly rendered this behaviour possible; they actually constitute it, and therefore implicitly endorse it.

Many think that justificatory neutrality is an unattainable and unattractive ideal even in the analogue city; I disagree, but nothing in my argument here depends on that. Justificatory neutrality is, other things equal, harder to achieve for algorithmic governance than for legal governance. If you agree that justificatory neutrality is an attractive ideal, this is a cause for concern; if you disagree, it remains a notable difference with which political philosophy must reckon. In my view the greater challenge of achieving justificatory neutrality in the Algorithmic City should again lead us to excavate the values that underpin it: it captures a desire for liberal pluralism grounded in concern for the values of liberty, equality, and collective self-determination described above. We should continue aiming for liberal pluralism, even if justificatory neutrality is beyond our reach. I pick up this thread in Lecture II of this book.

## 7. Objections

### 7.1. Review

I've introduced a model of how our social relations in the information age have changed, and have drawn out the distinctive modalities of algorithmic power that they entail. I've argued that this power should not simply be eliminated but must instead be justified against standards of procedural legitimacy, proper authority, and substantive justification. I then showed that, in applying these standards to the Algorithmic City, political philosophy must update how it thinks about each of them—our theories of authority should focus on the right to govern first, and the right to coerce derivatively; our theories of procedural legitimacy must expand to account for different means and modalities of governance, and must better foreground the importance of resistibility; our theories of substantive justification must face the obstacles that algorithmic governance interposes to the pursuit of justificatory neutrality, and aim for liberal pluralism instead.

In this section I focus on three objections that deflate the assumed challenge posed by algorithmic governance by arguing that we already have adequate resources to address it: first, because the stakes aren't high enough for the most demanding standards of political justification to apply to algorithmic intermediaries; second, because the Algorithmic City affords ease of exit, so we shouldn't worry if it is governed capriciously, or by those who lack authority to do so; third, because algorithmic intermediaries are, after all, a product of our extrinsically governed, physical world, and so can be subject to the regulatory power of states and state-like actors.

7.2. Stakes

The state can deprive you of your property, your freedom, even your life. The technologies involved in social media companies, messaging apps, search engines, and so on, exercise a very different kind of power. Surely the stringency of the justificatory standards that apply to algorithmic intermediaries should reflect the very different stakes?

The basic idea behind this objection is right—the procedural legitimacy and proper authority standards *should* be sensitive to the kind of power they are invoked to constrain. State power is different from the power of current algorithmic intermediaries. But I think the latter is still subject to exacting standards of legitimacy and authority. To show this, we need a theory of how the stakes of power affect the justificatory standards it must meet. What matters is not only power's measure but the purposes at which it aims.

First, it matters whether power is used to govern or for non-governance influence. Consider, for example, the difference between the pre-emptive prevention of copyright infringement and the delivery of targeted advertisements. Both involve the exercise of power by algorithmic intermediaries, but only the first counts as a species of governance because it aims to implement a norm of non-infringement of copyright. The second is a more naked exercise of influence, seeking to drive advertising sales. The reasons governing power demands procedural legitimacy and proper authority are explained at the end of section 4 above. One central and simple idea is that procedural legitimacy and proper authority are fundamental when A adjudicates disputes between B and C, thus shaping power relations between them.

Second, it matters whether power is used to pursue collective goods, such as the objects of group rights, or irreducibly social goods—whose enjoyment is conditional on knowing that others too enjoy it. Many goods at stake in the Algorithmic City are collective—for example, rights against data expropriation and exploitation that cannot adequately be understood through an individualist lens, or the value of civic robustness and a healthy information environment, as I argue in Lecture II.[140] If [A, B, C, and D] have a group right to X, and the contours of X are at least somewhat open to interpretation, then for A to act unilaterally to secure X might be defensible if no other alternative is available, but is pro tanto objectionable. It would be like a vanguard of revolutionaries effectively seceding from a parent state without having the support of people in the seceding territory.[141] Where irreducibly social goods and the objects of group rights are concerned, the members of the group at stake have a strong claim to exercise collective control over how those goods are realised. This increases the stakes for proper authority and likely for procedural legitimacy as a means to secure greater collective control of governing power.

---

[140] Benn and Lazar, 2021.
[141] A prospect grimly illustrated through the sham secessions of four regions of Ukraine, then annexed by Russia, in 2022.

As well as attending to the purposes of power, we must clearly attend to its stakes. But the stakes of power have different dimensions, each of which must be accounted for: degree, scope, and concentration.

The degree of power is a function of the interests at stake, separately and in the aggregate. The state's governing power is at one extreme—it can enforce its rules by confiscating property and freedom, conscripting its citizens, or taking their lives. Algorithmic intermediary governance often falls far short of this. Having one's Google account suspended may not be a trivial imposition, but falls far short of, for example, being conscripted and sent to war. And yet algorithmic intermediaries are central to realising some basic and essential individual interests in information and communication, which are crucial for our ability to do much else with our lives, and are likely protected by basic rights.

The degree of governing power is not simply a function of the welfare directly at stake for any particular individual. It is also a function of the individual interests indirectly at stake and the aggregate interests at stake for many individuals. For example, consider content moderation decisions by social media companies. Their direct impact on individuals will often be relatively low stakes, since they just amount to the removal of some post or other—though they can also be very significant (for example, when unmoderated online harassment drives someone to suicide). Their indirect impacts can be still more profound—for example, when algorithmic amplification helps to foment a genocidal furore or when subtle decisions about platform architecture contribute to the degeneration of the digital public sphere, with deleterious effects on democracy.[142] And sometimes the sheer scale of algorithmic intermediaries means that relatively trivial individual impacts amount, in the aggregate, to a significant impact overall and so a significant degree of power, as with some of the harms of targeted advertising.[143]

Second, scope concerns the range of people's choices that algorithmic intermediaries govern. Liberal democratic states generally limit the scope of government power, carving out spaces that are externally structured by regulation (specifying what the reasonable bounds of behaviour are), but are internally subject to their own norms. Algorithmic intermediaries exercise much more pervasive governing power. They are implicated in many domains that the state stays (more-or-less) out of, such as romantic, familial, and friendship relationships, as well as the discursive relationship between co-participants in the public sphere (my topic in Lecture II). Furthermore, they constitute the social relations that they mediate much more comprehensively than does the top-down governing power of the state: where the latter allows for unmediated interaction within the boundaries it sets, algorithmic intermediaries govern more pervasively, shaping everything possible by way of those intermediaries.

Third, the concentration of power identifies the ratio of those subject to a particular governing power to those controlling it. State power is highly concentrated;

---

[142] For a discussion of the role of Facebook in the Rohingya genocide in Myanmar, see Suzor, 2019 and https://www.amnesty.org/en/latest/news/2022/09/myanmar-facebooks-systems-promoted-violence-against-rohingya-meta-owes-reparations-new-report/.
[143] See Benn and Lazar, 2021 on 'stochastic manipulation', for example.

algorithmic governing power is often subject to the decisions of a single CEO. Of course, in the actual Algorithmic City (not Pre-Emptopolis) *many* companies are involved in any particular algorithmic intermediary. And yet the Algorithmic City's contours are largely determined by a handful of executives at the most powerful companies, with the potential for even more radical concentration of power when public companies are taken private.

The stakes of algorithmic intermediary governance are not, at this time, as high as for state governance. The measure of power is clearly less, and the state is more centrally concerned with the provision of collective goods and protection against the externalities of private decision-making. However, algorithmic intermediary governing power is nonetheless high stakes, and is on a path to play an ever bigger role in our lives. It would be a common mistake to exaggerate the governing power of algorithmic intermediaries, and put them on a par with states as they are now. But downplaying it would be an equal error, as would underestimating its potential. Recent progress in systems related to Large Language Models suggest that we may be at the precipice of a radical change in the degree to which algorithmic intermediaries pervade and govern our lives. Even if the next decade is spent only exploring and perfecting the economic, political, and cultural capabilities of existing multi-modal models like GPT-4o, without making any further significant advances, we are likely to see each of the major domains of human life be transformed by algorithmic intermediaries deploying generative AI. When Dialogue Agents provide the interface between people and government services, when universal intermediaries mediate everything that we do with digital technologies, curating the internet on our behalf and translating between us in all of our online interactions, when Generative AI Systems trained on proprietary data underpin data and investment management platforms that run our economies and our public services, the stakes of algorithmic intermediary power will be unquestionably on the kind of scale that demands robust standards of legitimacy and authority.

And of course, the next decade may signal even more substantial leaps forward than we have already experienced. The research labs responsible for the extraordinary progress in AI of the last ten years (and especially the last year) have the stated goal of bringing Artificial General Intelligence into existence—a technology by definition more powerful than anything humanity has ever created. Researchers at these labs often recognise the moral import of that undertaking, but they typically assume that their only task is to ensure, if they can, that these future AI systems are aligned with human values, or are 'provably beneficial'.[144] As alluded to above, in practice this means focusing narrowly on substantive justification—ensuring that these incredibly powerful future systems do only what we want them to do. But by showing the importance of procedural legitimacy and proper authority for justifying power, especially governing power, we can show their mistake.

This means, first, that aligning Generative AI models to 'human values' must clearly be about more than substantive justification. It also means attending closely to who gets to decide what values to align the models to, and ensuring that any power vested in these new systems is exercised legitimately. But, second, it may ultimately prove impossible to meet those standards (we don't even know how to ensure that they act with substantive justification; securing procedural legitimacy and proper authority is even more open a challenge). The most powerful AI systems are deeply, fundamentally inscrutable. They offer us no guarantees. It is very hard to see how they could satisfy even basic criteria of legitimacy, like a publicity requirement.[145] This should give us serious pause about whether we should even be trying to build AI systems of such unprecedented power. And this leads, third, to the most important question: on what basis should the AI labs consider themselves democratically authorised to pursue such a potentially disruptive technology? Observing procedural legitimacy and proper authority means not only ensuring that any highly-powerful AI system that the labs build is able to meet these exacting standards. It also means asking whether, as democratic polities, we want to be building such systems in the first place.

---

[144] Russell, 2019.

Algorithmic intermediaries already govern a significant fraction of our lives and are set to govern much more. It should be a priority to understand what it would take to realise proper authority and procedural legitimacy in the Algorithmic City.

7.3. Ease of Exit

As pervasive as the Algorithmic City is, and as much have we have come to depend on it, we are not (yet) bound to a single algorithmic intermediary in the same way as we are bound to a single state. Every platform has at least some competitors, and we can vote with our feet. Why isn't this enough to secure the authority of algorithmic governance and to mitigate any worries about procedural illegitimacy or substantive justification? If you don't like how this corner of the Algorithmic City is governed, you can simply leave—or set up your own competitor, governed according to principles you favour.

While a venerable tradition of political philosophy grounds the right to govern in the consent of the governed, there are some serious problems with this approach.[146] I will highlight three.[147]

First, the moral power of individual consent has limits—for example, when one's consent has significant externalities for others or when the object of one's consent inherently calls for collective, not individual, support. Consider a situation in which, should enough other people consent to X, my dissent would prove irrelevant, because the externalities of their choices mean that I am in the same position whether I consent or not. If I reasonably believe enough people will agree to X whatever I do, then my consent cannot be morally effective. It shows only that I would rather get whatever meagre benefits are associated with consenting to X than endure X without agreeing to it.[148] This point especially applies here—for example, I have argued elsewhere that our consent to share data with algorithmic intermediaries should be understood through this lens (thus resolving the 'privacy paradox').[149] In addition, we should think twice about appealing to individual consent when collective goods are at stake. If the function of governing power is to pursue irreducibly social goods or the objects of group rights, then its authority cannot plausibly derive just from individual consent, but should instead be grounded in support of the group indicated by those goods and rights.

---

[145] Lazar, 2024.

[146] Simmons, 1979. The objections to justifying state authority by appeal to consent go back a long way of course.

[147] The ease of exit objection persists despite many compelling demonstrations of its inadequacy. I aim to add to that literature in this subsection, but no doubt have missed some overlapping arguments. For some particularly compelling reasons to stop relying on exit to solve the internet's problems, see Barocas and Nissenbaum, 2014; Grimmelmann, 2014; Bietti, 2019. See also Slee, 2006.

[148] Barocas and Nissenbaum, 2014.

Second, the appeal to exit does not obviate the need to figure out how intermediary governance of the Algorithmic City could offer satisfactory answers to the 'how', 'who', and 'what' questions, rather it presupposes that we can come up with those answers. One's decision not to exit a regime only confers authority on its ruler if one has reasonable alternatives. For most of us, the costs of total exit from the Algorithmic City are too significant for our decision to remain to imply morally effective consent. But in that case, tacit consent must depend on our ability to exit one algorithmic regime in favour of another that can facilitate the same social relations. But a choice between awful regimes cannot confer authority on either. For your decision to remain in Algorithmic City A, given the availability of B, to confer authority on A, B must be a reasonable alternative. In this context, that surely means that B must meet standards of procedural legitimacy, proper authority, and substantive justification. So the tacit consent argument still relies on figuring out how algorithmic intermediaries could meet those standards.

Moreover, this argument for authority from tacit consent implies that sub-par algorithmic regimes can free-ride on the costs incurred by their morally conscientious counterparts: regime A may continue to govern you irresponsibly because you could have chosen responsible governance under B. A gets the benefit of participants/users (and all the value they bring with them) without the costs of governance. This is unjust on its face, all the more so given what we know about network effects in the Algorithmic City, which suggest that sub-par regimes might not only free-ride on responsible ones, but raise the costs of conscientious governance (TikTok springs to mind).

Third, the values of relational equality and collective self-determination offer further grounds for rejecting the appeal to ease of exit to justify governing power. The argument from exit presupposes that governing power's fundamental problem is how it undermines its subjects' negative liberty. But the moral hurdles faced by governing power concern relational equality and collective self-determination as much they do negative liberty. If A governs B, C, and D, this implicitly creates social hierarchy between them; and if A unilaterally determines the shared terms of their social existence, then B, C, and D are not collectively self-determining. Tacit consent is a fragile foundation for proper authority because it is consistent with A telling B, C, and D to 'take it or leave it'—scarcely an expression of egalitarian social relations—and with A calling all the shots and B, C and D having no opportunity for collective self-determination.

Rather than grounding authority in tacit consent derived from the ease of exit, I think that governing with proper authority when the stakes are relatively high requires allowing people to resist your rule. Electoral politics are part of this, but people must also be able to resist the very institutions by which you govern. If people can take opportunities for formal and informal resistance without high personal cost and decline to do so, then that *does* imply some endorsement of the governing power, which can help confer authority. It also implies a certain kind of relational equality—both on standard democratic grounds and because the possibility of resistance means those governing take those subject to power seriously enough to acknowledge that they might have got things wrong, and so recognise that there must always be extra-institutional means to challenge authority. And it indirectly favours collective self-determination, since it relies on our having direct influence on our community's direction of travel rather than only indirect influence through the threat of exit.

---

[149] Benn and Lazar, 2021.

As argued above, algorithmic intermediary power does not afford resistance. Algorithmic intermediaries need to be intentionally (re)designed to create this possibility. And to do this requires more than simply enabling people to move between different intermediaries, each with its own capricious and unauthorised governing power.

Finally, a note of caution on the recourse to competition as the solution for the Algorithmic City's ills. While ease of exit might not play a significant role in justifying governing authority, it *is* a fundamental obstacle to successful governance. Some degree of central authority is a precondition of effective governance.[150] This should provide a serious counterweight to attempts to resolve the pathologies of our current digital ecosystem through competition law and other cognate measures.

### 7.4. Governance of Algorithms?

The two key levers being deployed to resolve the pathologies of the Algorithmic City today are antitrust and regulation. I've just argued, in effect, that increasing competition is insufficient for legitimate and authoritative governance of the Algorithmic City. But couldn't we just regulate algorithmic intermediaries, specifying what count as legitimate procedures and appropriate substantive ends to aim at, grounding their authority to govern in their compliance with these regulations and thus with the will of the relevant jurisdiction?

I am obviously in favour of regulating algorithmic intermediaries. But my focus here is not on how we regulate the Algorithmic City but on governance within the Algorithmic City—how algorithmic intermediaries regulate us. Even if they are ultimately subject to the regulatory power of national or transnational governments, *we* are still subject to the power of algorithmic intermediaries. They shape power relations between us, and shape the impacts of those social relations on social structures over time. So we still need to answer each of the justificatory questions I articulated above. When we have those answers, we should determine which can be suitably enforced through regulation. But governing technology companies is different from figuring out how algorithmic intermediaries should govern us. And we should also be wary of rushing into regulations without first answering these more fundamental questions.[151]

---

[150] See Lecture II, and Benn and Lazar, 2021. For similar views, see Bietti, 2019.
[151] The European Union seems often to take more pride in being the world's first regulator of (western) digital technologies than it does in actually getting those regulations right. The Digital Services Act and the AI Act, show some promise, but they also force determinations on systems the capabilities of which we simply do not yet understand well enough for regulation to be successful.

This objection invites us to address a different question—of governance *of* algorithms, rather than governance *by* algorithms. But we can perhaps sharpen it, focusing on authority. Could one argue that regulation should dictate the procedural and substantive standards that apply to algorithmic intermediaries, and so any algorithmic intermediary that complies with these regulatory standards necessarily has proper authority to govern, as delegated by the regulating body?

The force of this argument depends, I think, on the relative importance of our time spent in the Algorithmic City and our analogue communities. The key question is whether A's authority over B, C, and D must be grounded in some facts about A, B, C, and D, or else can instead be delegated from some third-party institution, of which some subset of B–D are members. If B–D are not part of that third-party institution, then its authorisation seems morally irrelevant. Given the transnational nature of the Algorithmic City, this suggests that delegated state authority has limitations in the Algorithmic City. This is an obvious point but it bears repeating with additional emphasis: the Algorithmic City is fundamentally transnational, and for that reason poses deep challenges for the aspiration to democratic authority.

What's more, even if B–D *are* part of the broader community that authorises A, we might reject delegated authority, depending on how pervasive a role this algorithmic intermediary plays in B–D's lives.

Our reasons of relational equality and collective self-determination favour subjecting governing power to democratic authorisation, even when that governing power is itself subordinate to and regulated by a background democratic state. Or in other words, democratic states should not authorise private entities to engage in quasi-feudal governance of the Algorithmic City. Elizabeth Anderson reaches similar conclusions about the power of employers over employees—even though employers are authorised to act in this way by existing employment law, we still have reasons to oppose this kind of power; delegated authority is not an adequate replacement for democratic authorisation by the people being governed.[152]

The Algorithmic City as it exists today is poorly regulated. Better governance of algorithms is undoubtedly urgent. But we must still discover the principles that render algorithmic governance permissible. And as the Algorithmic City plays an ever greater role in our lives, the argument that its authority over us can be delegated from some other only partially-overlapping institution becomes ever weaker.

## 8. RECAP

Political philosophy should help us understand how to achieve free, egalitarian, collectively self-determining social relations. It can't do this if it is predicated on an outdated understanding of the nature of our social relations. In Lecture I of this book, I have sought to update political philosophy's understanding of social relations for the information age, focusing in particular on the advent of the Algorithmic City, the network of algorithmically mediated social relations made possible by computing, the internet, and more recently by advances in machine learning. I've argued that understanding and justifying algorithmic intermediary power is one of the central tasks for contemporary political philosophy. I've sought to illustrate just what algorithmic intermediary power is, and to introduce the tools analytic political philosophy offers to help justify the exercise of especially governing power—focusing on proper authority, procedural legitimacy, and substantive justification. I then showed how differences between algorithmic governance and especially extrinsic governance by law raise interesting questions for political philosophy with respect to each of these normative standards—and invite us to ask whether algorithmic governance might be in principle harder to justify than the modalities of power that it promises to supplant. Along the way I've shown that we should be especially sceptical of those who aspire to use algorithmic intermediaries to 'perfect' governance without attending to these risks, and of others who focus narrowly on 'aligning' algorithmic intermediaries with human values, where that invariably means only aspiring to substantive justification, rather than also emphasising the importance of procedural legitimacy and proper authority.

---

[152] Anderson, 2017.

The task ahead is vast: this Lecture offers only a descriptive and theoretical model, and a tasting menu of the key normative questions that must be addressed. In Lecture II, I explore these problems in greater depth for one particular corner of the Algorithmic City—the digital public sphere. But the challenges posed by algorithmic intermediaries will be untangled only through a profound and comprehensive practical reorientation of contemporary political philosophy. There is much work to do.